\documentclass[preprint,aps,prb,showpacs,amsmath,amssymb]{revtex4}
\usepackage{graphics}
\usepackage{epsf,graphicx,pstricks,epsfig,psfrag}
\begin{document}
\title{Microscopic approach to high-temperature superconductors: \\
Superconducting phase
}
\author{S.~Sykora and K.W.~Becker}

\affiliation{Institut f\"ur Theoretische Physik, Technische Universit\"at Dresden, 01062 Dresden, Germany}

\date{\today}



\begin{abstract}
Despite the intense theoretical and experimental effort, an understanding of the superconducting 
pairing mechanism of the high-temperature superconductors, leading to an unprecedented high transition
temperature $T_c$, is still lacking.
An additional puzzle is the unknown connection between the superconducting gap 
and the  so-called pseudogap which is a central property of the 
most unusual normal state.
 Starting from the $t$-$J$ model, 
we present a microscopic approach to the physical properties 
of the superconducting phase at moderate hole-doping in the framework of a novel renormalization scheme,
called PRM. This approach is based on a stepwise elimination of high-energy transitions using unitary
transformations. We arrive at a renormalized 'free'
Hamiltonian for the superconducting state. Our microscopic approach allows us
to explain the experimental findings in the underdoped 
as well as in the optimal hole doping regime. In good agreement with experiments, we
find no superconducting solutions for very small hole doping.
In the superconducting phase, the order parameter turns out to have $d$-wave 
symmetry with a coherence length of a few lattice constants. The spectral function, obtained
from angle-resolved photoemission spectroscopy (ARPES)
along the Fermi surface, is also in good agreement with experiment: The 
spectra display peak-like structures which are caused alone by coherent
excitations in a small range around the Fermi energy.
 
\end{abstract}


\maketitle

\section{Introduction}

Since the discovery of superconductivity in the cuprates \cite{BM86}, enormous
theoretical and experimental effort has been made to investigate 
the superconducting pairing mechanism which leads to an unprecedented high transition
temperature $T_c$\cite{C99}-\cite{V97}. The generic phase diagram of the cuprates shows a wide
variety of different behavior as a function of temperature and level of hole
doping. In particular, with increasing hole doping away from half-filling, 
the physical properties completely change at the transition to the superconducting phase.
 A large number of experiments using angle-resolved photoemission spectroscopy
(ARPES) have revealed a strong momentum dependence of the superconducting gap\cite{N98}-\cite{K08}.
An additional puzzle is the unknown connection 
between the superconducting gap of the superconducting phase 
and the so-called pseudogap which is a central 
property of the most unusual normal state of the cuprates. 

Superconductivity is usually understood
as an instability from a non-superconducting state. 
Therefore, often in  theoretical investigations, 
the starting point was either the Fermi-liquid  or the anti-ferromagnetic 
phase at large or low doping. In this paper, we take a different 
approach and only consider hole fillings, in which either a superconducting or a 
pseudogap phase is present. 
A generally accepted  model for the cuprates is the $t$-$J$ model which describes the electronic 
degrees of freedom  in the copper-oxide planes for low energies. Alternatively, one could also start
from a one-band Hubbard Hamiltonian as a minimal model. 
However, for low energy excitations, the latter model reduces to the $t$-$J$ model, 
so that both models are equivalent. In a preceding paper \cite{I}, henceforth
denoted by I,  we have investigated the pseudogap phase 
in the cuprates on the basis of the $t$-$J$ model. Our aim is to extent the
microscopic approach from paper~I to the superconducting phase.
As our theoretical tool, we use a recently developed 
projector-based renormalization method which is called PRM \cite{PRM}.
The approach is based on a stepwise elimination of high-energy transitions using unitary
transformations. We thus arrive at a renormalized 'free'
Hamiltonian for correlated electrons which can describe both the
superconducting phase and the pseudogap phase. 
For the superconducting phase, the order parameter turns out to have $d$-wave 
symmetry with a coherence length of a few lattice constants. The basic feature for the understanding of the superconducting pairing mechanism 
in the underdoped regime is a characteristic electronic oscillation behavior between neighboring
lattice sites. The oscillation becomes less important for larger $\delta$
which agrees with the weakening of the superconducting phase for larger hole
doping. The spectral function, obtained
from angle-resolved photoemission spectroscopy (ARPES)
along the Fermi surface, also agrees well with experiment: The 
spectra display peak-like structures which are caused alone by coherent
excitations in a small range around the Fermi energy.

After a short introduction of the model in Sec.~II, we apply the projector-based 
renormalization method (PRM) in Sec.~III to the $t$-$J$ model. 
The results will be discussed in Sec.~IV.

\section{Model}

In the preceding paper I,  
we have investigated the pseudogap phase 
in the cuprates on the basis of the $t$-$J$ model. 
We adopt the same model also for the superconducting phase of the hole-doped cuprates. 
As before, we restrict ourselves to moderate hole concentrations away from half-filling
outside the antiferromagnetic phase
 \begin{eqnarray}
\label{1}
{\cal H} &=& -\sum_{ij, \sigma} t_{ij}\, \hat c_{i\sigma}^\dagger \hat c_{j\sigma} 
- \mu \sum_{i\sigma} \hat c_{i\sigma}^\dagger \hat c_{i\sigma} 
+ \sum_{ij} J_{ij} {\bf S}_i {\bf S}_j    
=: {\cal H}_t + {\cal H}_J .
\end{eqnarray}
The $t$-$J$ Hamiltonian 
consists of a conditional hopping term and an antiferromagnetic exchange interaction
and acts in a unitary space with empty and singly occupied sites.
The Hubbard creation and annihilation operators $\hat {c}_{i\sigma}^{(\dagger)}
= {c}_{i\sigma}^{(\dagger)} (1- n_{i,-\sigma})$ in Eq.~\eqref{1}
obey nontrivial anti-commutator relations 
\begin{eqnarray}
\label{3}
[\hat c_{i\sigma}^\dagger, \hat c_{j\sigma'}]_+ &=& \delta_{ij} \big( \delta_{\sigma \sigma'}
{\cal D}_\sigma(i) + \delta_{\sigma, -\sigma'} S_{i}^{\sigma} \big) .
\end{eqnarray}
Here, ${\bf S}_{\bf q}$ is the local spin operator and ${\cal D}_\sigma (i)$ is defined by
${\cal D}_\sigma (i) = 1 - n_{i,-\sigma}$.
In Fourier notation,  the $t$-$J$ model \eqref{1} reads
\begin{eqnarray}
\label{6}
 {\cal H} &=&  \sum_{{\bf k}, \sigma} \varepsilon_{\bf k}\, \hat c_{{\bf k}\sigma}^\dagger 
\hat c_{{\bf k}\sigma} 
+ \sum_{\bf k} \left(\Delta_{{\bf k},\Lambda} \hat c_{{\bf k}\uparrow}^\dagger 
\hat c_{{\bf k}\downarrow}^\dagger +
 \Delta_{{\bf k},\Lambda}^* \hat c_{{\bf k}\downarrow} \hat c_{{\bf k}\uparrow} \right) 
+
\sum_{{\bf q}} J_{{\bf q}} {\bf S}_{\bf q} {\bf S}_{-{\bf q}} .
\end{eqnarray}
$\varepsilon_{\bf k}$  measures the one-particle energy from the Fermi energy 
$\varepsilon_{\bf k} =- \sum_{i\neq (j)} t_{ij} e^{i{\bf k}({\bf R}_i -{\bf R}_j)} - \mu$.
 Note that in Eq.~\eqref{6}, we have introduced an infinitesimal field
$\Delta_{{\bf k},\Lambda} \rightarrow 0$ 
which breaks the gauge symmetry in the superconducting phase.

\section{Renormalization approach for the superconducting Phase}
\label{Re-app}

Let us apply the PRM to the $t$-$J$ model in the superconducting phase. 
We consider the case of  moderate hole-doping, where   
superconductivity occurs. As before, the 
hopping element $t$ between nearest neighbors is assumed to be large compared 
to the exchange coupling $J$. Therefore, we can decompose the Hamiltonian  
into an 'unperturbed' part ${\cal H}_0$ and into a 'perturbation' ${\cal H}_1$, 
\begin{eqnarray}
\label{28}
 {\cal H}_0 &=&  
\sum_{{\bf k}\sigma} \varepsilon_{\bf k} \, \hat c_{{\bf k}\sigma}^\dagger  \hat c_{{\bf k}\sigma} 
+\sum_{\bf k} \left(\Delta_{{\bf k},\Lambda} \hat c_{{\bf k}\uparrow}^\dagger 
\hat c_{{\bf k}\downarrow}^\dagger +
 \Delta_{{\bf k},\Lambda}^* \hat c_{{\bf k}\downarrow} \hat c_{{\bf k}\uparrow} \right) 
+\sum_{\bf q} J_{\bf q}\, {\cal A}_0({\bf q}), \nonumber\\
{\cal H}_1 &=& \sum_{\bf q} J_{\bf q}\, \left( {\cal A}_1({\bf q}) + {\cal A}_1^\dagger({\bf q})
\right).
\end{eqnarray} 
The decomposition \eqref{28} is an extension of the former decomposition for the pseudogap phase 
to the superconducting phase. It is based on a splitting of the exchange into two
parts. The first one, containing ${\cal A}_0$, commutes with ${\cal H}_t$ and should, therefore, be
a part of the unperturbed Hamiltonian ${\cal H}_0$. In contrast, the two operators  ${\cal A}_1$
and ${\cal A}_1^\dagger$ do not commute with ${\cal H}_t$ 
and belong to ${\cal H}_1$. They are defined by
\begin{eqnarray}
 \label{26}
{\cal A}_0({\bf q}) &=& \frac{1}{2}\left( {\bf S}_{{\bf q}}{\bf S}_{-{\bf q}} +
\frac{1}{\hat\omega_{\bf q}^2} \dot{\bf S}_{{\bf q}}\dot{\bf S}_{-{\bf q}} 
\right),
\\
{\cal A}_1({\bf q}) &=& \frac{1}{4}\left({\bf S}_{\bf q} - \frac{i}{\hat \omega_{\bf q}}\,
 \dot{\bf S}_{\bf q}\right)
\, \left({\bf S}_{-{\bf q}} - \frac{i}{\hat \omega_{\bf q}}\, \dot{\bf S}_{-{\bf q}}\right), \nonumber
\\
{\cal A}_1^\dagger({\bf q})  &=& 
\frac{1}{4}\left({\bf S}_{\bf q} + \frac{i}{\hat \omega_{\bf q}}\,
 \dot{\bf S}_{\bf q}\right)
\, \left({\bf S}_{-{\bf q}} + \frac{i}{\hat \omega_{\bf q}}\, \dot{\bf S}_{-{\bf q}}\right), 
\nonumber
\end{eqnarray} 
and obey approximately the following relations:
\begin{eqnarray}
\label{27}
{\sf L}_0\,{\cal A}_0({\bf q}) &=&  0, \qquad 
{\sf L}_0\, {\cal A}_1({\bf q}) = 2 \hat \omega_{\bf q}\,  {\cal A}_1({\bf q}),   \qquad
{\sf L}_0\,{\cal A}_1^\dagger({\bf q})  = -2\hat \omega_{\bf q}\, {\cal A}_1^\dagger({\bf q})  .
\end{eqnarray}
Here, ${\sf L}_0$ is the Liouville operator corresponding to ${\cal H}_0$,
where ${\sf L}_0$ is defined by ${\sf L}_0 {\cal C}
= [{\cal H}_0, {\cal C}] $ for any operator variable ${\cal C}$,  and $\hat \omega_{\bf q}$ is
given by 
\begin{eqnarray}
\label{22}
\hat \omega^2_{\bf q} &=&
  2P_0(t^2_{{\bf q}=0}- t^2_{{\bf q}}) =  \hat \omega^2_{-{\bf q}} \geq 0, \quad \qquad
t^2_{\bf q} = \sum_{l (\neq i)}t_{il}^2\, e^{i{\bf q}({\bf R}_l - {\bf R}_i)} \, .
\end{eqnarray}

\subsection{Renormalization equations}
The derivation of the renormalization equations for the parameters of the Hamiltonian 
runs parallel to that for the pseudogap phase.
The aim of the projector-based renormalization method (PRM) is to eliminate all transitions 
due to  ${\cal H}_1$ between the eigenstates of ${\cal H}_0$ with non-zero transition energies.
Let us assume that all excitations with energies larger than a given cutoff $\lambda$ 
have already been eliminated. Then, an {\it ansatz} for  
the renormalized Hamiltonian ${\cal H}_\lambda$ should have the following form,
\begin{eqnarray}
 {\cal H}_\lambda = {\cal H}_{0,\lambda} +{\cal H}_{1,\lambda} 
\label{29}
\end{eqnarray}
with
\begin{eqnarray}
\label{29a}
 {\cal H}_{0,\lambda} &=&   {\cal H}_{t,\lambda}
+\sum_{\bf q} J_{{\bf q},\lambda}\, {\cal A}_{0,\lambda}({\bf q})  
- \sum_{\bf k} \left( \Delta_{{\bf k},\lambda}\, \hat c_{{\bf k}, \uparrow}^\dagger 
\hat c_{-{\bf k}, \downarrow}^\dagger + \Delta_{{\bf k},\lambda}^*\, 
\hat c_{-{\bf k}, \downarrow} \hat c_{{\bf k}, \uparrow}
\right)   + E_\lambda   ,    \\
{\cal H}_{1,\lambda} &=& \sum_{\bf q} J_{{\bf q},\lambda}\, \Theta(\lambda -|2\hat \omega_{{\bf q}, \lambda}|)
 \left( {\cal A}_{1,\lambda} ({\bf q}) + {\cal A}_{1,\lambda}^\dagger({\bf q})
\right)  \, . \nonumber
\end{eqnarray}  
$ {\cal H}_{t,\lambda} =
\sum_{{\bf k}\sigma} \varepsilon_{{\bf k},\lambda} \, \hat c_{{\bf k}\sigma}^\dagger  
\hat c_{{\bf k}\sigma} $ is the renormalized hopping term and depends on $\lambda$. 
The other parameters
$\Delta_{{\bf k},\lambda}$, $\hat \omega_{{\bf q},\lambda}$,
and $J_{{\bf q},\lambda}$ in Eq.~\eqref{29a} are also $\lambda$-dependent. However, the 
$\lambda$-dependence of $J_{{\bf q},\lambda}$ can  be suppressed according to paper~I.

The $\lambda$-dependent operators ${\cal A}_{\alpha,\lambda}({\bf q})$ ($\alpha=0, \pm 1$) in Eqs.~\eqref{29a} 
are defined as in Eqs.~\eqref{26}. However, 
$\dot {\bf S}_{\bf q}$  and $\hat \omega_{{\bf q}}$ have to be 
replaced by $\dot {\bf S}_{{\bf q},\lambda}$ and $\hat \omega_{{\bf q},\lambda}$,
\begin{eqnarray}
\dot {\bf S}_{{\bf q},\lambda} &=& \frac{i}{\hbar} [{\cal H}_{0,\lambda}, {\bf S}_{{\bf q},\lambda}] 
\approx \frac{i}{\hbar} [{\cal H}_{t,\lambda}, \omega_{\bf q}],  \\
\hat\omega_{{\bf q},\lambda}^2 &=& 2P_0\,(t^2_{{\bf q}=0, \lambda} -
  t^2_{{\bf q}, \lambda})\, , \qquad
t^2_{{\bf q},\lambda} = \sum_{i(\neq j)}t^2_{{ij},\lambda}\, e^{i{\bf q}({\bf
  R}_i- {\bf R}_j)}. \nonumber
\end{eqnarray}

In order to derive renormalization equations for the parameters of ${\cal H}_\lambda$, 
we eliminate all excitations within an additional energy shell between 
$\lambda$ and a  
reduced cutoff $\lambda - \Delta \lambda$. According to paper~I, this is done by
applying a unitary transformation to ${\cal H}_\lambda$,
\begin{eqnarray}
  \label{12}
  \mathcal{H}_{(\lambda - \Delta\lambda)} &=& 
  e^{X_{\lambda,\Delta\lambda}} \, \mathcal{H}_{\lambda} \,
  e^{-X_{\lambda,\Delta\lambda}}\, .
\end{eqnarray}

The generator $X_{\lambda, \Delta \lambda}$ was constructed in paper I and is given
in lowest order perturbation theory by Eq.~(I.37), 
\begin{eqnarray}
\label{31}
 X_{\lambda, \Delta \lambda} &=& \sum_{\bf q}\frac{J_{{\bf q}}}{2 \hat \omega_{{\bf q}, \lambda}}
\Theta_{\bf q}(\lambda, \Delta\lambda)\left({\cal A}_{1,\lambda}({\bf q}) - 
{\cal A}_{1,\lambda}^\dagger ({\bf q})  \right) .
\end{eqnarray}
Here, $\Theta_{\bf q}(\lambda, \Delta\lambda)$ denotes a product of two $\Theta$-functions
\begin{eqnarray*}
\Theta_{\bf q}(\lambda, \Delta\lambda) &=& \Theta(\lambda -|2 \hat \omega_{{\bf q},\lambda}|)\,
\Theta\left (|2\omega_{{\bf q}, \lambda -\Delta \lambda}| -(\lambda -\Delta \lambda) \right),
\end{eqnarray*}
which confines the elimination range to excitations with 
$|2\omega_{{\bf q}, \lambda -\Delta \lambda}|$ larger than $\lambda - \Delta \lambda$
and $|2 \hat \omega_{{\bf q},\lambda}|$ smaller than $\lambda$.
Roughly speaking, for the  case of a weak $\lambda$-dependence of 
$|\omega_{{\bf q},\lambda}|$,
the elimination is restricted to all transitions within the energy  shell 
between $\lambda -\Delta \lambda$ and $\lambda$. 
According to Eqs.~\eqref{26}, the generator $X_{\lambda, \Delta \lambda}$ can also be expressed by
\begin{eqnarray}
\label{32}
 X_{\lambda, \Delta \lambda} &=& 
-i \sum_{\bf q}\frac{J_{{\bf q}}}{ 4\hat \omega_{{\bf q}, \lambda}^2}
\Theta_{\bf q}(\lambda, \Delta\lambda)\left(
{\bf S}_{\bf q} \,\dot {\bf S}_{-{\bf q},\lambda} + \dot {\bf S}_{{\bf q},\lambda} \, {\bf S}_{-{\bf q}} 
 \right) .
\end{eqnarray}

The explicit evaluation of  the unitary transformation \eqref{12} follows that of paper I. In perturbation theory 
to second order in $J_{\bf q}$, one finds
\begin{eqnarray}
\label{33}
{\cal H}_{\lambda - \Delta \lambda} &=& e^{X_{\lambda, \Delta \lambda}}\,
{\cal H}_{\lambda} \, e^{-X_{\lambda, \Delta \lambda}} = 
{\cal H}_{\lambda - \Delta \lambda}^{(0)} + {\cal H}_{\lambda - \Delta \lambda}^{(1)} 
+{\cal H}_{\lambda - \Delta \lambda}^{(2)}  + \cdots,
\end{eqnarray}
where 
\begin{eqnarray}
\label{34}
{\cal H}_{\lambda - \Delta \lambda}^{(0)} &=& 
{\cal H}_{t,\lambda}
- \sum_{\bf k} \left( \Delta_{{\bf k},\lambda}\, \hat c_{{\bf k}, \uparrow}^\dagger 
\hat c_{-{\bf k}, \downarrow}^\dagger + \Delta_{{\bf k},\lambda}^*\, 
\hat c_{-{\bf k}, \downarrow} \hat c_{{\bf k}, \uparrow}
\right) + E_\lambda ,   \nonumber \\
{\cal H}_{\lambda - \Delta \lambda}^{(1)} &=&
\sum_{\bf q} J_{{\bf q}}\, {\cal A}_{0,\lambda}({\bf q}) 
+ [X_{\lambda, \Delta \lambda}, {\cal H}_{t, \lambda}]
+ 
 \sum_{\bf q} J_{{\bf q}}\, \Theta(\lambda -|2\hat \omega_{{\bf q}, \lambda}|)
 \left( {\cal A}_{1,\lambda} ({\bf q}) + {\cal A}_{1,\lambda}^\dagger({\bf q})
\right),
\nonumber \\
%
{\cal H}_{\lambda - \Delta \lambda}^{(2)} &=&
\frac{1}{2} [X_{\lambda, \Delta \lambda},  [X_{\lambda, \Delta \lambda}, {\cal H}_{t, \lambda}]\, ]
+
\sum_{\bf q} J_{{\bf q}} \,  [X_{\lambda, \Delta \lambda}, {\cal A}_{0,\lambda}({\bf q})] \nonumber \\
&& +
\sum_{\bf q} J_{{\bf q}}\, \Theta(\lambda -|2\hat \omega_{{\bf q}, \lambda}|)\,
 [ \, X_{\lambda, \Delta \lambda},
{\cal A}_{1,\lambda} ({\bf q}) + {\cal A}_{1,\lambda}^\dagger({\bf q}) \,].
\end{eqnarray}
All expressions agree with those of paper~I, except that in ${\cal H}_{\lambda - \Delta \lambda}^{(0)}$
the new symmetry breaking terms appear.  The commutators can be evaluated as in paper I.
Let us at first investigate the effect of the second order term 
${\cal H}_{\lambda - \Delta \lambda}^{(2)}$.  The obtained operator expressions have to be reduced in a 
further factorization approximation to operator terms appearing 
in ${\cal H}_\lambda$. Thereby, also a reduction 
to operators  $\hat c_{{\bf k}\uparrow}^\dagger
\hat c_{-{\bf k}\downarrow}^\dagger$ and $\hat c_{-{\bf k}\downarrow} \hat c_{{\bf k}\uparrow} $
has to be included.  The final result has to be compared with the  formal expression 
for ${\cal H}_{\lambda- \Delta \lambda}$, which corresponds to the expression \eqref{29}
for ${\cal H}_\lambda$, when $\lambda$ is replaced by $\lambda - \Delta \lambda$.  
According to Appendix A, the following  second order renormalizations to 
$\varepsilon_{{\bf k},\lambda}$ and to the order parameter $\Delta_{{\bf k},\lambda}$ 
are found
\begin{eqnarray}
\label{35}
 \varepsilon_{{\bf k},\lambda - \Delta \lambda}- \varepsilon_{{\bf k},\lambda} &=&
\frac{1}{16 N}\sum_{\bf q} \frac{J_{\bf q}^2}{\hat \omega_{{\bf q}, \lambda}^4}\,
\Theta_{\bf q}(\lambda, \Delta \lambda) \,
(\varepsilon_{{\bf k}+ {\bf q}, \lambda} +  \varepsilon_{{\bf k}- {\bf q}, \lambda}
-2 \varepsilon_{{\bf k}, \lambda} )\, \langle \dot {\bf S}_{{\bf q},\lambda} \, 
\dot {\bf S}_{-{\bf q},\lambda} \rangle \nonumber \\
&& + 
\frac{3}{2N} \sum_{{\bf q}\sigma} \left(\frac{J_{\bf q}}{4 \hat \omega_{{\bf q},\lambda}^2}\right)^2 \,
\Theta_{\bf q}(\lambda,\Delta \lambda) \,(\varepsilon_{{\bf k},\lambda}-
\varepsilon_{{\bf k}- {\bf q},\lambda })^2 \\
&& \times \left[ \frac{1}{N} \sum_{{\bf k}'\sigma'}(2\varepsilon_{{\bf k}',\lambda} -
\varepsilon_{{\bf k}'+{\bf q},\lambda} -\varepsilon_{{\bf k}'-{\bf q},\lambda}) 
\langle \hat c_{{\bf k}'\sigma'}^\dagger  \hat c_{{\bf k}'\sigma'} \rangle 
\right] \, n_{{\bf k}-{\bf q}\alpha}^{(NL)} ,
  \nonumber \\
 \Delta_{{\bf k},\lambda - \Delta \lambda}- \Delta_{{\bf k},\lambda} &=&
-\frac{1}{16 N}\sum_{\bf q} \frac{J_{\bf q}^2}{\hat \omega_{{\bf q}, \lambda}^4}\,
\Theta_{\bf q}(\lambda, \Delta \lambda) \,
(\varepsilon_{{\bf k}, \lambda} -\varepsilon_{{\bf k}+{\bf q}, \lambda})^2 
\langle \hat c_{-({\bf k} +{\bf q})\downarrow}  \hat c_{{\bf k} +{\bf q}\uparrow} \rangle \nonumber \\  
&& \times\frac{1}{N} \sum_{{\bf k}'} 
(\varepsilon_{{\bf k}'+ {\bf q}, \lambda} +  \varepsilon_{{\bf k}'- {\bf q}, \lambda}
-2 \varepsilon_{{\bf k}', \lambda} )\,  n_{{\bf k}'\sigma}^{(NL)} ,
\label{36}
\end{eqnarray}
where we have defined
\begin{eqnarray}
\label{35a}
 n_{{\bf k},\sigma}^{(NL)}  &=&
\langle \hat c_{{\bf k}\sigma}^\dagger \hat c_{{\bf k}\sigma} \rangle 
- \frac{1}{N} \sum_{{\bf k}'}\langle \hat c_{{\bf k}'\sigma}^\dagger \hat
c_{{\bf k}'\sigma} \rangle  
\end{eqnarray}
as non-local part of the one-particle occupation number per spin direction.
An equivalent equation also exists for $E_{\lambda -\Delta \lambda}$.
The quantity $\langle \dot {\bf S}_{{\bf q},\lambda}  \dot {\bf S}_{-{\bf
    q},\lambda} \rangle$ is a correlation function of the time derivatives of
${\bf S}_{{\bf q}}$  and was evaluated in paper~I.
Note that an additional contribution to $\varepsilon_{{\bf k}, \lambda - \Delta \lambda}$, proportional to 
the correlation function $\langle {\bf S}_{\bf q}\cdot {\bf S}_{-{\bf q}} \rangle$, 
has already been neglected. 
The remaining expectation values in \eqref{35}, \eqref{36} have to be calculated separately. In principle, 
they should be defined with the $\lambda$-dependent Hamiltonian ${\cal H}_\lambda$, because the factorization approximation was employed for the renormalization step from ${\cal H}_\lambda$ to ${\cal H}_{\lambda -
\Delta \lambda}$.  However, ${\cal H}_\lambda$  still contains interactions which prevent a 
straight evaluation of $\lambda$-dependent expectation values.
The  best way to circumvent this difficulty is to calculate the expectation values with the full Hamiltonian ${\cal H}$ instead of with   ${\cal H}_\lambda$. In this case, 
the renormalization equations can be solved self-consistently, as it was done in paper~I.

Up to now, the renormalization contributions were evaluated from the
second order term ${\cal H}_{\lambda- \Delta \lambda}^{(2)}$  of  ${\cal H}_{\lambda- \Delta \lambda}$. 
Inserting $\varepsilon_{{\bf k},\lambda -\Delta \lambda}$ and 
$\Delta_{{\bf k},\lambda -\Delta \lambda}$  into Eq.~\eqref{33}, we obtain
\begin{eqnarray} 
\label{37}  
{\cal H}_{\lambda -\Delta \lambda} &=&  
{\cal H}_{t,\lambda- \Delta \lambda}
- \sum_{\bf k} \left( \Delta_{{\bf k},\lambda - \Delta \lambda}\, \hat c_{{\bf k}, \uparrow}^\dagger 
\hat c_{-{\bf k}, \downarrow}^\dagger + \Delta_{{\bf k},\lambda- \Delta \lambda}^*\, 
\hat c_{-{\bf k}, \downarrow} \hat c_{{\bf k}, \uparrow}
\right) +  {\cal H}_{\lambda -\Delta \lambda}^{(1)} + E_{\lambda - \Delta \lambda}.
\end{eqnarray}
The first order term ${\cal H}_{\lambda -\Delta \lambda}^{(1)}$ has still to be evaluated.
This can be done along the procedure of paper I. The final result for the renormalized Hamiltonian 
${\cal H}_{\lambda- \Delta \lambda}$ reads
${\cal H}_{\lambda - \Delta 
\lambda}= {\cal H}_{0,\lambda - \Delta \lambda}
+ {\cal H}_{1, \lambda - \Delta \lambda}$,  with 
\begin{eqnarray}
 \label{41}
{\cal H}_{0,\lambda - \Delta \lambda} &=&
{\cal H}_{t,\lambda- \Delta \lambda}
- \sum_{\bf k} \left( \Delta_{{\bf k},\lambda - \Delta \lambda}\, \hat c_{{\bf k}, \uparrow}^\dagger 
\hat c_{-{\bf k}, \downarrow}^\dagger + \Delta_{{\bf k},\lambda- \Delta \lambda}^*\, 
\hat c_{-{\bf k}, \downarrow} \hat c_{{\bf k}, \uparrow}
\right) + E_{\lambda - \Delta \lambda}  \\
 && +
\sum_{\bf q} J_{{\bf q}}\, {\cal A}_{0,\lambda- \Delta \lambda}({\bf q}), \nonumber \\
{\cal H}_{1, \lambda -\Delta \lambda} &=& 
\sum_{\bf q} J_{{\bf q}}\,
\Theta(\lambda -\Delta \lambda -|\hat \omega_{{\bf q}, \lambda - \Delta \lambda}|)\, 
\left({\cal A}_{1,\lambda - \Delta \lambda}({\bf q}) +{\cal A}_{1,\lambda- \Delta \lambda}^\dagger
({\bf q})\right). \nonumber
\end{eqnarray}
The renormalized Hamiltonian ${\cal H}_{\lambda - \Delta \lambda}$ 
has the same operator structure as
${\cal H}_\lambda$. Therefore, we can formulate a renormalization procedure as follows:
We start from the original $t$-$J$ model in the presence of a small gauge symmetry breaking 
field. 
The energy cutoff of the original model is denoted by  
${\lambda= \Lambda}$. Starting from a guess for the unknown expectation values,
which enter the renormalization equations \eqref{35} and \eqref{36}, 
we proceed by eliminating all  excitations in steps $\Delta \lambda$ 
from $\lambda=\Lambda$ down to $\lambda=0$.
Thereby, the parameters of the Hamiltonian change in steps according to the
renormalization equations \eqref{35} and \eqref{36}. In this way, we obtain a final model at $\lambda =0$, 
in which the perturbation  ${\cal H}_{1,\lambda}$  is completely integrated out.
It reads
\begin{eqnarray}
 \label{42}
{\cal H}_{\lambda =0} &=& \sum_{{\bf k}\sigma} \varepsilon_{{\bf k},\lambda=0}\,
\hat c_{{\bf k} \sigma}^\dagger \, \hat c_{{\bf k} \sigma}
- \sum_{\bf k} \left( {\Delta}_{{\bf k},\lambda=0}\, \hat c_{{\bf k}, \uparrow}^\dagger 
\hat c_{-{\bf k}, \downarrow}^\dagger + {\Delta}_{{\bf k},\lambda=0}^*\, 
\hat c_{-{\bf k}, \downarrow} \hat c_{{\bf k}, \uparrow}
\right)    \nonumber \\
 &+&  \sum_{\bf q} J_{{\bf q}}\, {\cal A}_{{0},\lambda=0}({\bf q}) + E_{\lambda=0}     .
\end{eqnarray}
Unfortunately, due to the presence of the ${\cal A}_0$-term, 
the result \eqref{42} does not yet allow us
to recalculate the expectation values, since the eigenvalue problem of ${\cal H}_{\lambda =0}$
can not be solved. 
Therefore, a further approximation is necessary. It consists of
a factorization of the second term in 
\begin{eqnarray}
\label{43}
 \sum_{\bf q} J_{{\bf q}}\, {\cal A}_{{0},\lambda=0}({\bf q}) &=& 
\sum_{\bf q}
\frac{J_{\bf q}}{2}\left( {\bf S}_{{\bf q}}{\bf S}_{-{\bf q}} +
\frac{1}{\hat \omega_{{\bf q},\lambda=0}^2} 
\dot{\bf S}_{{\bf q}, \lambda=0}\dot{\bf S}_{-{\bf q},\lambda=0} \right) .
\end{eqnarray}
According to Appendix A, we end up with a modified Hamiltonian which will be denoted by
$\tilde{\cal H}^{(1)}$, 
\begin{eqnarray}
 \label{44}
\tilde {\cal H}^{(1)} &=& \sum_{{\bf k}\sigma} \tilde \varepsilon_{\bf k}^{(1)}\,
\hat c_{{\bf k} \sigma}^\dagger \, \hat c_{{\bf k} \sigma}
- \sum_{\bf k} \left( \tilde {\Delta}_{{\bf k}}^{(1)}\, \hat c_{{\bf k}, \uparrow}^\dagger 
\hat c_{-{\bf k}, \downarrow}^\dagger + \tilde {\Delta}_{{\bf k}}^{(1)*}\, 
\hat c_{-{\bf k}, \downarrow} \hat c_{{\bf k}, \uparrow}
\right)    
 +\sum_{\bf q} \frac{J_{{\bf q}}}{2}\, {\bf S}_{\bf q}\,  {\bf S}_{-{\bf q}} 
 + \tilde E^{(1)} .      \nonumber \\
&& 
\end{eqnarray}
Here, not only the electron energy $\varepsilon_{{\bf k},\lambda=0}$ but also the order parameter $\Delta_{{\bf k},\lambda=0}$
is modified according to
\begin{eqnarray}
\label{45}
\tilde \varepsilon_{\bf k}^{(1)} &=& 
\varepsilon_{{\bf k},\lambda=0} - \frac{1}{N} \sum_{\bf q} \frac{3J_{\bf q}}
{4 \hat \omega^2_{{\bf q},\lambda=0}} (\varepsilon_{{\bf k},\lambda=0} - 
\varepsilon_{{\bf k}+ {\bf q},\lambda=0} )^2\, n_{{\bf k} +{\bf
    q},\sigma}^{(NL)}, \nonumber \\
\tilde {\Delta}_{{\bf k}}^{(1)} &=& {\Delta}_{{\bf k},\lambda=0} 
-\frac{1}{N} \sum_{\bf q} \frac{3J_{\bf q}}
{4 \hat \omega^2_{{\bf q},\lambda=0}} (\varepsilon_{{\bf k},\lambda=0} - 
\varepsilon_{{\bf k}+ {\bf q},\lambda=0} )^2\, 
\langle \hat c_{-({\bf k}+{\bf q})\downarrow}\, 
\hat c_{{\bf k}+{\bf q}\uparrow}\, 
\rangle  ,
\end{eqnarray} 
where $n_{{\bf k}\sigma}^{(NL)}$ is defined in Eq.~\eqref{35a}. 
Note that the operator structure of $\tilde {\cal H}^{(1)}$ agrees 
with that of the original $t$-$J$ model of Eq.~\eqref{6} in the presence of
the symmetry breaking field. 
However, the parameters have changed. Most important, the strength of the 
exchange coupling in Eq.~\eqref{44} is decreased by a factor of $1/2$. This property allows us to start 
the whole renormalization procedure again. We consider 
the modified $t$-$J$ model \eqref{44} as our new initial Hamiltonian (at $\lambda= \Lambda$)
which again has to be renormalized. The initial values of the new Hamiltonian 
$\tilde{\cal H}^{(1)}$ at cutoff $\lambda =\Lambda$ are
$\tilde \varepsilon_{\bf k}^{(1)}$, $\tilde {\Delta}_{{\bf k}}^{(1)}$, 
and $J_{\bf q}/2$. After the new renormalization cycle, the exchange coupling of the 
renormalized Hamiltonian $\tilde{\cal H}^{(2)}$ is again decreased by a factor of $1/2$,
until, after a sufficiently large number of renormalization cycles 
($n\rightarrow \infty$), the exchange completely disappears. 
Thus, we finally arrive at a 'free' model
\begin{eqnarray}
 \label{46}
\tilde {\cal H} &=& \sum_{{\bf k}\sigma} \tilde \varepsilon_{\bf k}\,
\hat c_{{\bf k} \sigma}^\dagger \, \hat c_{{\bf k} \sigma}
- \sum_{\bf k} \left( \tilde {\Delta}_{{\bf k}}\, \hat c_{{\bf k}, \uparrow}^\dagger 
\hat c_{-{\bf k}, \downarrow}^\dagger + \tilde {\Delta}_{{\bf k}}^{*}\, 
\hat c_{-{\bf k}, \downarrow} \hat c_{{\bf k}, \uparrow}
\right)     + \tilde E \, .
\end{eqnarray}
Here, we have introduced the new notation,
$\tilde {\cal H} =  \tilde {\cal H}^{(n\rightarrow \infty)}$, $\tilde \varepsilon_{\bf k} =
\tilde \varepsilon_{\bf k}^{(n \rightarrow \infty)}$, 
$\tilde \Delta_{\bf k}= \tilde {\Delta}_{{\bf k}}^{(n\rightarrow \infty)}$,
and $\tilde E = \tilde E^{(n \rightarrow \infty)}$.
Note that the Hamiltonian $\tilde {\cal H}$ 
allows us  to recalculate the unknown expectation values. With these 
 values, the whole renormalization procedure can be 
started again, until, after a sufficiently large number of such overall cycles, 
the expectation values converge. Then, 
the renormalization equations have been solved self-consistently.
However, the fully renormalized Hamiltonian \eqref{46} is actually not a 'free' model.
Instead, it is still subject to strong electronic correlations which are built in  
by the presence of the Hubbard operators.

\subsection{Evaluation of expectation values}
\label{expvalues}
The expectation values in Eqs.~\eqref{35}, \eqref{36}, and \eqref{45} are formed with the full Hamiltonian.  
To evaluate an expectation value 
$\langle {\cal A}\rangle$, we have to apply the unitary transformation also on 
the operator variable ${\cal A}$,
\begin{eqnarray}
\label{47}
 \langle {\cal A}\rangle &=& \frac{\mbox {Tr}\,({\cal A}\, e^{-\beta{\cal H}})}
{\mbox{Tr}\,e^{-\beta{\cal H}}} =
 \langle {\cal A}(\lambda) \rangle_{{\cal H}_\lambda} = 
\langle \tilde{\cal A} \rangle_{\tilde{\cal H}} \, ,
 \end{eqnarray}
where we have defined $
{\cal A}(\lambda) = e^{X_{\lambda}} \; {\cal A} e^{-X_{\lambda}} $ and  
${\tilde{\cal A}} = {\cal A}(\lambda \rightarrow 0)
$. Thus, additional renormalization equations for ${\cal A}(\lambda)$
have to be derived. 

\subsubsection{ARPES spectral functions}
First, let us consider the spectral function from 
angle resolved photoemission (ARPES). It is defined by
\begin{eqnarray}
\label{48}
 { A}({\bf k}, \omega)  &=& \frac{1}{2\pi} \int_{-\infty}^{\infty}
\big< \hat c_{{\bf k}\sigma}^\dagger (-t)  \;\hat c_{{\bf k}\sigma} 
\big> \; e^{i\omega t} dt = 
\big< \hat c_{{\bf k}\sigma }^\dagger \, \delta( {\sf L} + \omega ) \;
\hat c_{{\bf k}\sigma} \big>
\end{eqnarray}
and can be rewritten by use of the dissipation-fluctuation theorem as
\begin{eqnarray}
\label{49}
 { A}({\bf k}, \omega)  &=&   \frac{1}{1+ e^{\beta \omega}} \Im  G({\bf k}, \omega) \, .
\end{eqnarray}
Here, $\Im G({\bf k},\omega)$ is the dissipative part of the anti-commutator Green 
function,
\begin{eqnarray*}
\Im G({\bf k}, \omega) &=& \frac{1}{2\pi} \, \int_{-\infty}^{\infty}
\big< [\hat c_{{\bf k}\sigma}^\dagger (-t)\, ,  \;\hat c_{{\bf k}\sigma}]_+ 
\big> \; e^{i\omega t} dt =
\big< [\hat c_{{\bf k}\sigma }^\dagger \, , \, \delta( {\sf L} + \omega ) \;
\hat c_{{\bf k}\sigma}]_+ \big>. \nonumber 
\end{eqnarray*}
The time dependence and the expectation value are formed 
with the full Hamiltonian ${\cal H}$, and $\sf L$ is the Liouville operator corresponding to 
${\cal H}$.
According to Eq.~\eqref{47},  the anti-commutator Green function can be expressed by
\begin{eqnarray} 
\label{50}
 {\Im  G}({\bf k}, \omega)  &=&  
\big< [\hat c_{{\bf k}\sigma }^\dagger(\lambda) \, ,\, \delta ({\sf L}_\lambda + \omega) \;
\hat c_{{\bf k}\sigma}(\lambda)]_+ \big>_\lambda \, ,
\end{eqnarray}
where the creation and annihilation operators are subject 
to the unitary transformation. In order 
to derive renormalization equations for $\hat c_{{\bf k}\sigma}(\lambda)$ and 
$\hat c_{{\bf k}\sigma}^\dagger(\lambda)$, we restrict ourselves 
to a weak coupling theory. In this case, all contributions to the unitary transformation 
from the symmetry breaking fields can be neglected. 
Therefore, we can take over the previous {\it ansatz} (I.59) 
for  $\hat c_{{\bf k}\sigma}(\lambda)$ from paper I:
\begin{eqnarray}
\label{51}
 \hat c_{{\bf k} \sigma}(\lambda) &=& u_{{\bf k}, \lambda} \hat c_{{\bf k}\sigma} +
\frac{1}{2N}\sum_{{\bf q k}'} v_{{\bf k}, {\bf q},\lambda} \, \frac{J_{\bf q}}{4 
\hat \omega^2_{{\bf q},\lambda}} \, 
 \sum_{\alpha \beta \gamma} (\vec \sigma_{\alpha \beta}
\cdot \vec \sigma_{\sigma \gamma}) 
 (\varepsilon_{{\bf k}',\lambda}- \varepsilon_{{\bf k}' + {\bf q},\lambda }) \, 
 \,
 \hat c^\dagger_ {{\bf k}' + {\bf q} \alpha} \ \hat c_ {{\bf k}' \beta} \
\hat c_ {{\bf k} + {\bf q} \gamma} . \nonumber \\
&&
\end{eqnarray} 
Note that 
the dominant $\lambda$-dependence of $\hat c_{{\bf k} \sigma}(\lambda)$ 
is transfered to the 
parameters $u_{{\bf k},\lambda}$ and $v_{{\bf k},{\bf q},\lambda}$.
The general renormalization scheme was already established in paper I. 
Thus, running through the renormalization cycle many times ($n \rightarrow \infty$),
the exchange interaction will completely be eliminated. 
For $n\rightarrow\infty$, we arrive at the fully renormalized operator 
\begin{eqnarray}
\label{57a}
 \hat c_{{\bf k} \sigma}^{(n\rightarrow \infty)}(\lambda=0) &=& \tilde u_{{\bf k}} 
\hat c_{{\bf k}\sigma} +
\frac{1}{2N}\sum_{{\bf q k}'} \tilde{v}_{{\bf k}, {\bf q}} \,
 \frac{J_{\bf q}}{4 \tilde{\omega}^2_{\bf q}} \, 
 \sum_{\alpha \beta \gamma} (\vec \sigma_{\alpha \beta}
\cdot \vec \sigma_{\sigma \gamma}) 
(\tilde{\varepsilon}_{{\bf k}'}- \tilde{\varepsilon}_{{\bf k}' + {\bf q}}) \, 
\hat c^\dagger_ {{\bf k}' + {\bf q} \alpha} \ \hat c_ {{\bf k}' \beta} \
\hat c_ {{\bf k} + {\bf q} \gamma}\, ,  \nonumber \\
&&
\end{eqnarray} 
where $ \tilde u_{\bf k}=
u_{{\bf k},\lambda = 0}^{(n \rightarrow \infty)}$,
$ \tilde{v}_{{\bf k},{\bf q}}=
v_{{\bf k}, {\bf q}, \lambda = 0}^{(n \rightarrow \infty)}$, and $\tilde{\varepsilon}_{\bf k}=
\varepsilon_{{\bf k},\lambda=0}^{(n \rightarrow \infty)}$. Using the renormalized Hamiltonian 
$\tilde{\cal H}$ of Eq.~\eqref{46}, the spectral function 
$\Im G({\bf k},\omega)$ can be transformed to
\begin{eqnarray}
\label{86a}
 {\Im  G}({\bf k}, \omega)  &=&  
\big< [\hat c_{{\bf k}\sigma }^{(n\rightarrow \infty)\dagger}(\lambda = 0) \, ,\, \delta (\tilde{\sf L} + \omega) \;
\hat c_{{\bf k}\sigma}^{(n\rightarrow \infty)}(\lambda=0)]_+ \big>_{\tilde{\cal H}},
\end{eqnarray}
where the Liouville operator  $\tilde{\sf L}$ is related to $\tilde{\cal H}$.  
The expectation value has to be evaluated with $\tilde{\cal H}$. For this purpose, we
introduce new approximate quasiparticle operators (Appendix~B),
\begin{eqnarray}
\label{87a}
\alpha_{\bf k}^\dag &=& {\sf U}_{\bf k}\, \hat c_{{\bf k},\uparrow}^\dag - {\sf V}_{\bf k}\,
\hat c_{-{\bf k},\downarrow}, \nonumber \\
\beta_{\bf k}^\dag &=& {\sf U}_{\bf k}\, \hat c_{-{\bf k},\downarrow}^\dag + {\sf V}_{\bf k}\,
\hat c_{{\bf k},\uparrow}  , 
\end{eqnarray}
which fulfill the following relations:  $\tilde{\sf L}
  \alpha_{\bf k}^\dag = E_{\bf k} \alpha_{\bf k}^\dag$ and $\tilde{\sf L}
  \beta_{\bf k}^\dag = E_{\bf k} \beta_{\bf k}^\dag$, where $E_{\bf k} = \sqrt{ \tilde 
\varepsilon_{\bf k}^2 + D^2 \tilde \Delta_{\bf k}^2}$.
Inserting Eq.~\eqref{57a} into Eq.~\eqref{86a} and replacing all $c_{\bf k\sigma}^{(\dagger)}$-operators by  
the quasiparticle operators $\alpha_{\bf k}^{(\dagger)}$ and $\beta_{\bf k}^{(\dagger)}$, the $\delta$-functions 
can be evaluated. For the expectation values,  we restrict ourselves to the
leading order in the superconducting order parameter. The resulting
expression for $\Im G({\bf k},\omega)$ reads:
\begin{eqnarray}
\label{88}
 {\Im  G}({\bf k}, \omega) &=& \frac{D \tilde u_{\bf k}^2}{2} \left\{ \left( 1 + \frac{\tilde \varepsilon_{\bf
 k}}{E_{\bf k}} \right) \delta\left(\omega - E_{\bf k}\right) + \left( 1 - \frac{ \tilde \varepsilon_{\bf
 k}}{E_{\bf k}} \right) \delta\left(\omega + E_{\bf k}\right) \right\}  \\
&+& 
\frac{3D}{2N^2} \sum_{{\bf q}{\bf q}'}
\left[
\left( \frac{J_{\bf q}\tilde v_{{\bf k},{\bf q}}}{4 \hat \omega_{\bf q}^2}  \right)^2
 ( \varepsilon_{{\bf k}+{\bf q}'} - \varepsilon_{{\bf k}+{\bf q}+{\bf q}'} 
)^2  \right. \nonumber \\
&&\times
\left\{ \tilde n_{{\bf k}+{\bf q}+{\bf q}'} \tilde m_{{\bf k}+{\bf q}'} +
 \tilde n_{{\bf k}+{\bf q}} (D+ \tilde n_{{\bf k}+{\bf q}'} 
- \tilde n_{{\bf k}+{\bf q}+{\bf q}'} ) \right\} 
 \nonumber \\
&-& \frac{1}{2}\frac{J_{\bf q}}{4 \hat \omega_{\bf q}^2}
\frac{J_{{\bf q}'}}{4 \hat \omega_{{\bf q}'}^2} \,
\tilde v_{{\bf k},{\bf q}}\, \tilde v_{{\bf k},{\bf q}'}\,  
( \varepsilon_{{\bf k}+{\bf q}'} - \varepsilon_{{\bf k}+{\bf q}+{\bf q}'} )
( \varepsilon_{{\bf k}+{\bf q}} - \varepsilon_{{\bf k}+{\bf q}+{\bf q}'} ) \nonumber \\
&& \times \left\{ (\tilde n_{{\bf k}+{\bf q}'} -  \tilde m_{{\bf k}+{\bf q}})
\tilde  n_{{\bf k}+{\bf q}+{\bf q}'} 
- \tilde n_{{\bf k}+{\bf q}'} ( \tilde n_{{\bf k}+{\bf q}} + D)  \right\}
 \Bigg]  \nonumber \\
&&  \times \delta \left\{ \omega + \mbox{sign}(\tilde \varepsilon_{{\bf
 k}+{\bf q}+{\bf q}'}) E_{{\bf k}+{\bf q}+{\bf q}'} -
\mbox{sign}(\tilde \varepsilon_{{\bf k}+{\bf q}'}) E_{{\bf k}+{\bf q}'} -
 \mbox{sign}(\tilde \varepsilon_{{\bf k}+{\bf q}}) E_{{\bf k}+{\bf q}} 
\right\} \, , \nonumber
\end{eqnarray}
where $\tilde{n}_{\bf k}$ and $\tilde{m}_{\bf k}$ are defined by
 $\tilde{n}_{\bf k}= \langle \hat c_{{\bf k}\sigma}^\dagger \hat c_{{\bf k}\sigma}\rangle_{\tilde{\cal H}} $ 
and  $\tilde{m}_{\bf k}= \langle \hat c_{{\bf k}\sigma} \hat c_{{\bf k}\sigma}^\dagger\rangle_{\tilde{\cal H}}$. 
 For $\tilde n_{\bf k}$ and $\tilde m_{\bf k}$, we use the 
Gutzwiller approximation\cite{GW},  
 \begin{eqnarray}
\label{60}
 \tilde n_{{\bf k}} &=& (D-q)  + q  \, f(\tilde \varepsilon_{\bf k}) ,  \\
\tilde m_{\bf k} &=& q  \, (1- f(\tilde \varepsilon_{\bf k}))  \quad 
\mbox{with} \quad q = \frac{1-n}{1-n/2} \, ,
\nonumber
\end{eqnarray}
where $f(\tilde \varepsilon_{\bf k})$ is the 
Fermi function,  $f(\tilde \varepsilon_{\bf k})= 
\Theta( - \tilde{\varepsilon}_{\bf k})$ for $T=0$. 
Note that 
$\tilde m_{{\bf k} }$ is proportional to the hole 
filling $\delta = 1 - n$. Obviously, the application of $\hat c_{{\bf k}\sigma}^\dagger$
on a Hilbert space vector is non-zero only when holes are present. 
In contrast,  $\tilde n_{{\bf k} \sigma}$ does not vanish even at half-filling.

\subsubsection{Pair correlation function} 
In order to evaluate the superconducting order parameter $\tilde \Delta_{\bf k}$,
we have to know the superconducting pairing function
$\langle \hat c_{-{\bf k}\downarrow} \, \hat c_{{\bf k}\uparrow} \rangle$. Here, the 
expectation value is defined  with the full Hamiltonian for the superconducting phase. 
We first have to transform the pairing function, according to Eq.~\eqref{47} 
\begin{eqnarray*}
\langle \hat c_{-{\bf k}\downarrow} \, \hat c_{{\bf k}\uparrow} \rangle &=&
\langle \hat c_{-{\bf k}\downarrow}(\lambda) \, \hat c_{{\bf k}\uparrow}(\lambda)
 \rangle_{{\cal H}_\lambda} \, ,
\end{eqnarray*}
where the expectation value is now formed with the Hamiltonian 
${\cal H}_\lambda$,  given by Eq.~\eqref{29}.
In a weak coupling theory, all contributions from the 
symmetry breaking fields to the unitary transformation of $\hat c_{-{\bf k}\downarrow}(\lambda)$ and 
$\hat c_{{\bf k}\uparrow}(\lambda)$ can again be neglected. Therefore, we can immediately take over our previous 
result \eqref{51} for $\hat c_{{\bf k},\sigma}(\lambda)$. 
For the full renormalization ($n \rightarrow \infty$), we obtain
\begin{eqnarray}
\label{64}
 \langle \hat c_{-{\bf k}\downarrow} \, \hat c_{{\bf k}\uparrow} \rangle &=&
\tilde u_{\bf k}^2 \, 
\langle \hat c_{-{\bf k}\downarrow} \, \hat c_{{\bf k}\uparrow}
\rangle_{\tilde{\cal H}} \\
&+& \frac{3}{2N^2} \sum_{{\bf q}{\bf k}'} \tilde v_{{\bf k}, {\bf q}}^2 \, \left(\frac{J_{\bf q}}{4 \hat \omega_{\bf q}^2}\right)^2
(\varepsilon_{{\bf k}'} - \varepsilon_{{\bf k}'+{\bf q}})^2 \tilde m_{{\bf k}'+{\bf q}} \tilde n_{{\bf k}'} 
\langle \hat c_{-({\bf k}+{\bf q})\downarrow} 
\hat c_{({\bf k}+{\bf q})\uparrow} \rangle_{\tilde {\cal H}}   \nonumber \, .
\end{eqnarray}
Contributions
from third order in the superconducting order parameter have been neglected. The expectation values on the 
right hand side are formed with  
the fully renormalized Hamiltonian $\tilde{\cal H}$ (Eq.~\eqref{46}).  Using again 
the approximate Bogoliubov transformation of Appendix~\ref{C}, we find
 \begin{eqnarray}
 \label{66}
\langle \hat c_{-{\bf k}\downarrow} \, \hat c_{{\bf k}\uparrow}
 \rangle_{\tilde{\cal H}} &=& \frac{D^2 \tilde \Delta_{\bf
  k}}{2 E_{\bf k}} \left( 1 - \frac{2}{1 + e^{\beta E_{\bf k}}} \right)  .
\end{eqnarray}


\section{Numerical evaluation for the superconducting state}

Superconducting solutions have been obtained by evaluating self-consistently 
the full PRM renormalization scheme for a sufficiently large number of renormalization cycles. We
have taken the same parameters as for the normal state in subsection~V~B of
paper I, $t' = 0.4t$, $J = 0.2t$.

\subsection{Order parameter}
\begin{figure}
  \begin{center}
    \scalebox{1.0}{
      \includegraphics*{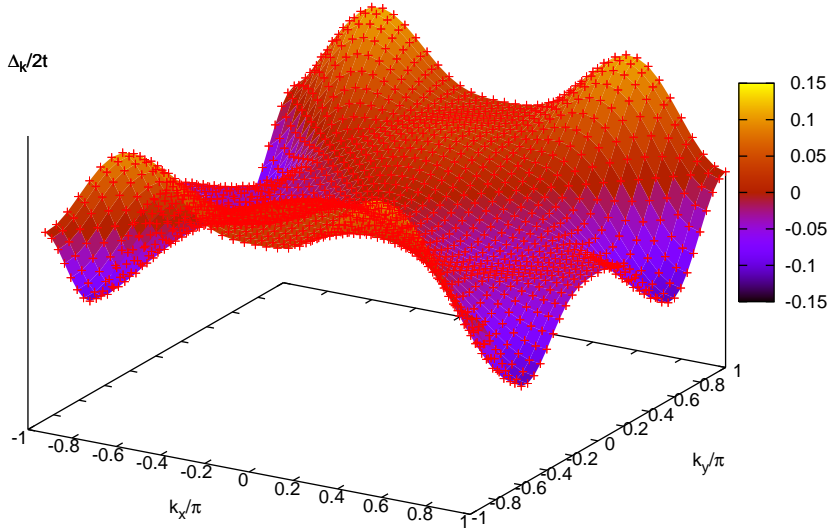} 
    }
  \end{center}
\caption{The superconducting gap function $\tilde\Delta_{\bf k}$ versus ${\bf k}$,
as obtained from Eq.~\eqref{69} for a square lattice with $N=40 \times 40$ sites. The parameters are 
$\delta = 0.08$, $t' = 0.4t$, $T=0$. 
Note that the gap function shows $d$-wave symmetry.  
}
\label{Fig_9}
\end{figure}

\begin{figure}
  \begin{center}
    \scalebox{0.95}{
      \hspace{-2.5cm}\includegraphics*{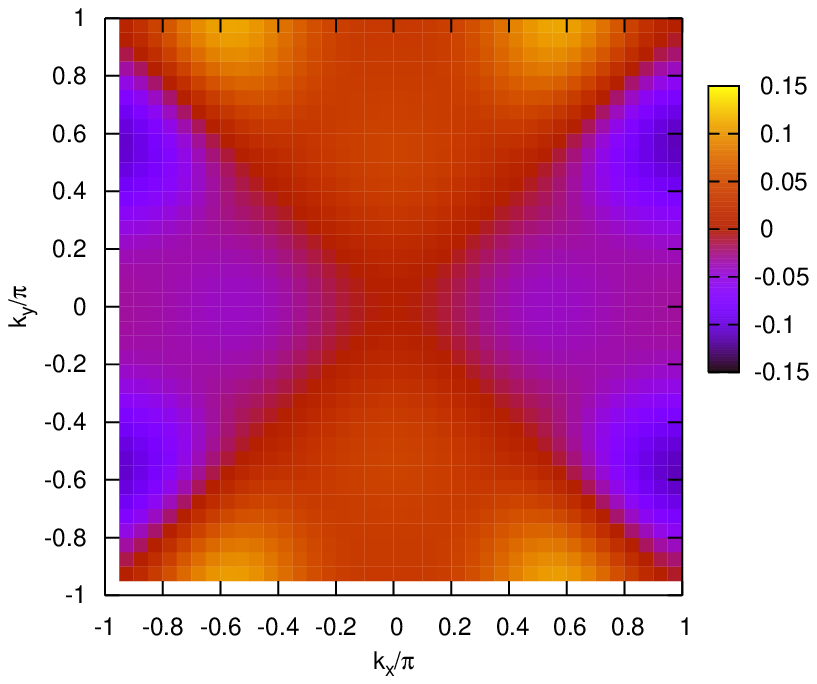} \hspace{-4cm}
      \includegraphics*{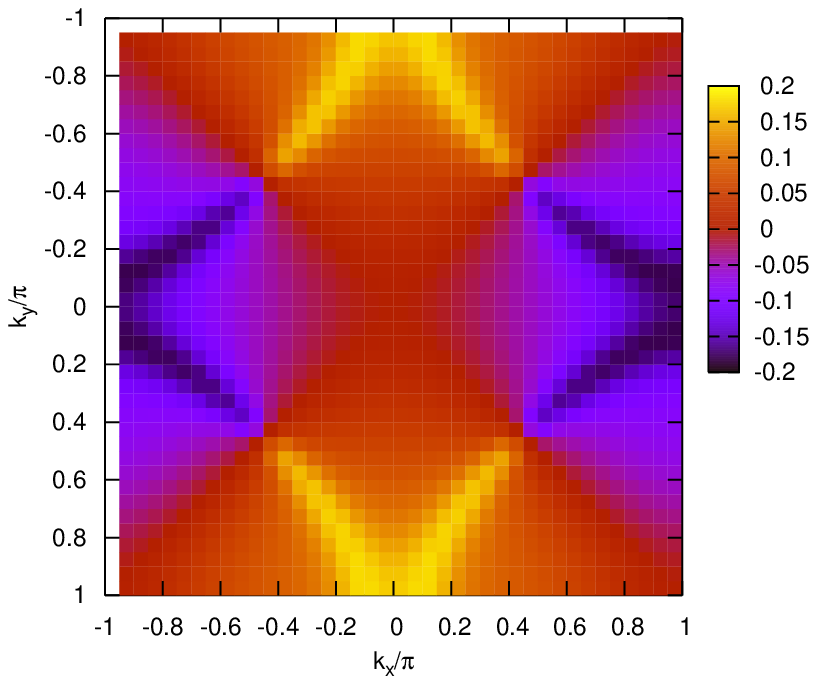} 
    }
  \end{center}
\caption{The superconducting gap $\tilde\Delta_{\bf k}$ (left
      panel) and the superconducting pairing function $\langle \hat c_{-{\bf k}\downarrow}\, 
\hat c_{{\bf k}\uparrow} \rangle$
for the same parameters as in Fig.~\ref{Fig_9} plotted as a 2d map.
}
\label{Fig_10}
\end{figure}

\subsubsection{Zero temperature results} 
In Fig.~\ref{Fig_9}, the superconducting gap function $\tilde\Delta_{\bf k}$ 
is plotted in ${\bf k}$-space for optimal doping, $\delta = 0.08$. 
In agreement with experiment, the solution shows $d$-wave symmetry
with nodal lines directed along the diagonals of the Brillouin zone from 
$(-\pi , -\pi)$ to $(\pi , \pi)$  and from
$(\pi , -\pi)$ to $(-\pi , \pi)$.
No $s$-wave like solutions were found. 

In Fig.~\ref{Fig_10}, both the superconduction gap function  $\tilde\Delta_{\bf k}$ (left panel) and
the pair correlation function $\langle \hat c_{-{\bf k}\downarrow}\, 
\hat c_{{\bf k}\uparrow} \rangle$ (right panel) are shown as a 2d-plot for the same
parameter values as in Fig.~\ref{Fig_9}. Again, in both functions, the nodal lines 
are clearly seen. Moreover, the absolute value of the pair correlation
$|\langle \hat c_{-{\bf k}\downarrow}\, \hat c_{{\bf k}\uparrow} \rangle|$ 
has a pronounced maximum along the Fermi surface (FS).
This behavior can easily be understood from 
Eq.~\eqref{66}. For ${\bf k}$-values close to the FS, ${\bf k} \approx {\bf k}_F$, 
where $\varepsilon_{\bf k} \leq {\cal O}(\tilde\Delta_{\bf k})$, the quantity 
$|\langle \hat c_{-{\bf k}\downarrow}\, \hat c_{{\bf k}\uparrow} \rangle|$ 
is of order  ${\cal O}(1)$. In contrast, for ${\bf k}$-vectors away from the FS (with
$\varepsilon_{\bf k} \gg {\cal O}( \tilde\Delta_{\bf k})$),
the pair correlation function  is of order ${\cal O}(\Delta / t)$. Note that the gap function 
$|\tilde\Delta_{\bf k}|$ has only a weak minimum at the Fermi surface. 
Additional weak maxima can be detected for the 
following ${\bf k}$-vectors: $(\pm\pi,\pm0.55\pi)$, $(\pm0.55\pi,\pm\pi)$, $(\pm0.5\pi,0)$ and $(0,\pm0.5\pi)$. 

\begin{figure}
  \begin{center}
    \scalebox{0.61}{
      \includegraphics*{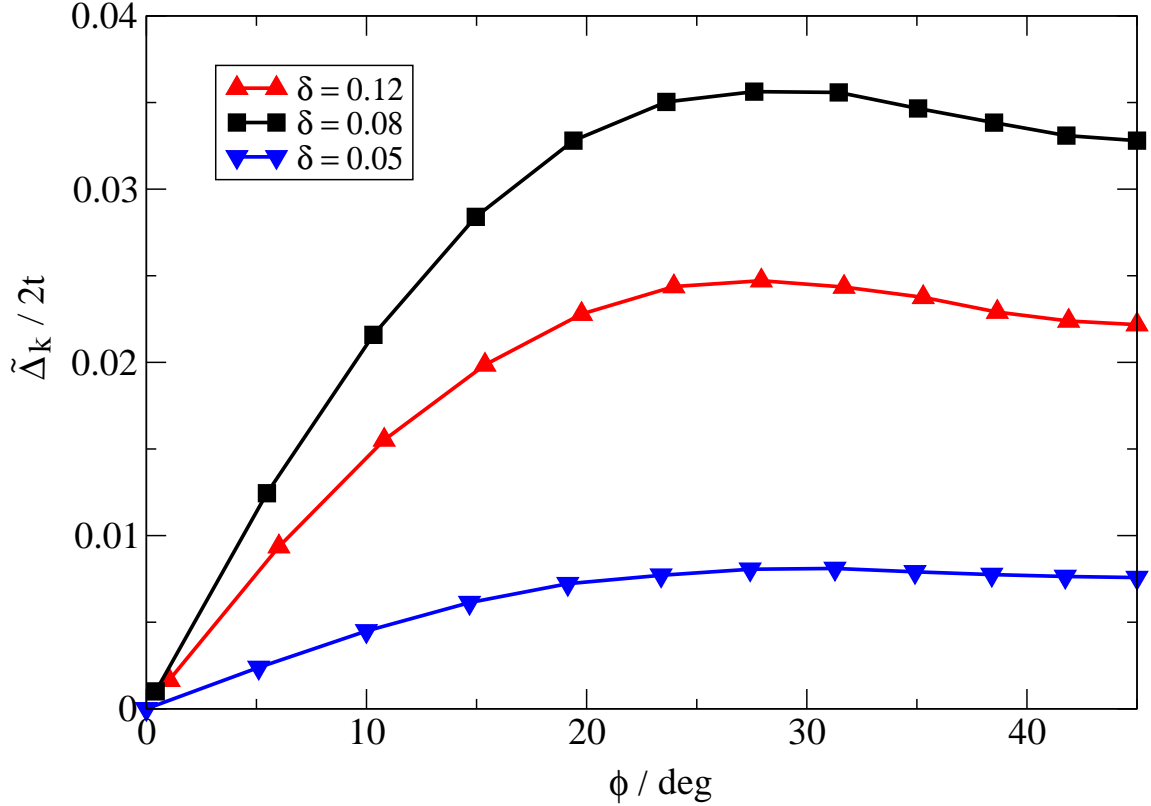}
    }
  \end{center}
  \caption{Superconducting gap function $\tilde \Delta_{\bf k}$ (in units of $2t$)
      as a function of the Fermi surface angle $\phi$ which was defined in the inset of
      Fig.~3 of paper~I for three doping values, $\delta = 0.05$
      (underdoped case, blue line), $\delta = 0.08$ (optimal doping, black line),
      and $\delta = 0.12$ (overdoped case, red line). }
\label{Fig_11}
\end{figure}

Fig.~\ref{Fig_11} shows the superconducting gap function $\tilde\Delta_{\bf k}$  
on the Fermi surface as a function of the Fermi surface angle $\phi$
for three doping values, $\delta = 0.05$
(underdoped case, blue line), $\delta = 0.08$ (optimally doped, black line),
and $\delta = 0.12$ (overdoped, red line). The angle $\phi$ was  already
 defined in paper I in the inset of Fig.~3.
 In all three cases, $\tilde\Delta_{\bf k}$ 
 shows a characteristic overall increase from the nodal ($\phi=0$) 
to the anti-nodal point. Note, however, that the  maximum value is
  already reached at a finite angle of about $27^\circ$, which is followed by a weak
decrease of $\tilde\Delta_{\bf k}$.


\begin{figure}
  \begin{center}
    \scalebox{0.71}{
      \includegraphics*{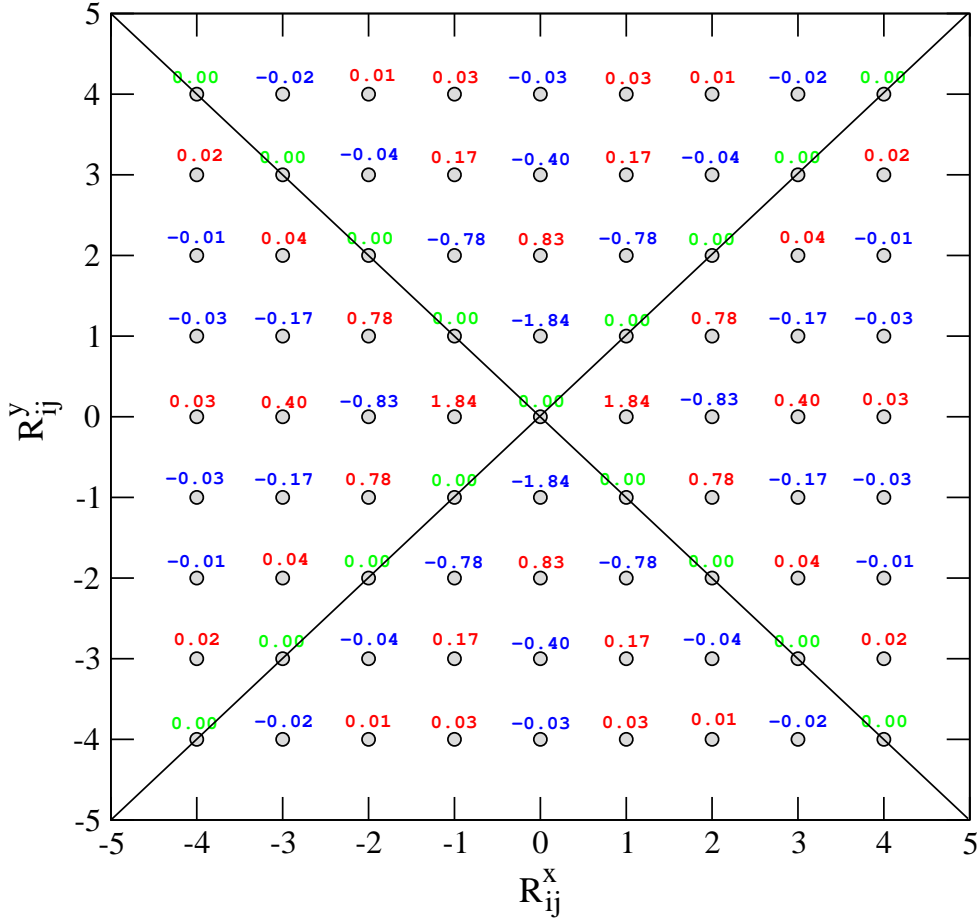}
    }
  \end{center}
  \caption{Superconducting order parameter in local space $\tilde \Delta_{ij}$ 
(in units of $10^{-2}(2t)$) for optimal doping $\delta=0.08$ and $T=0$. 
The hopping parameter $t'$ between next-nearest neighbors is given by $t'= 0.4t$. 
$R_{ij}^x$ and $R_{ij}^y$ denote the $x$ and $y$ components of
${\bf R}_i -  {\bf R}_j$.}
\label{Fig_12}
\end{figure}

\begin{figure}
 \begin{center}
    \scalebox{0.61}{
      \includegraphics*{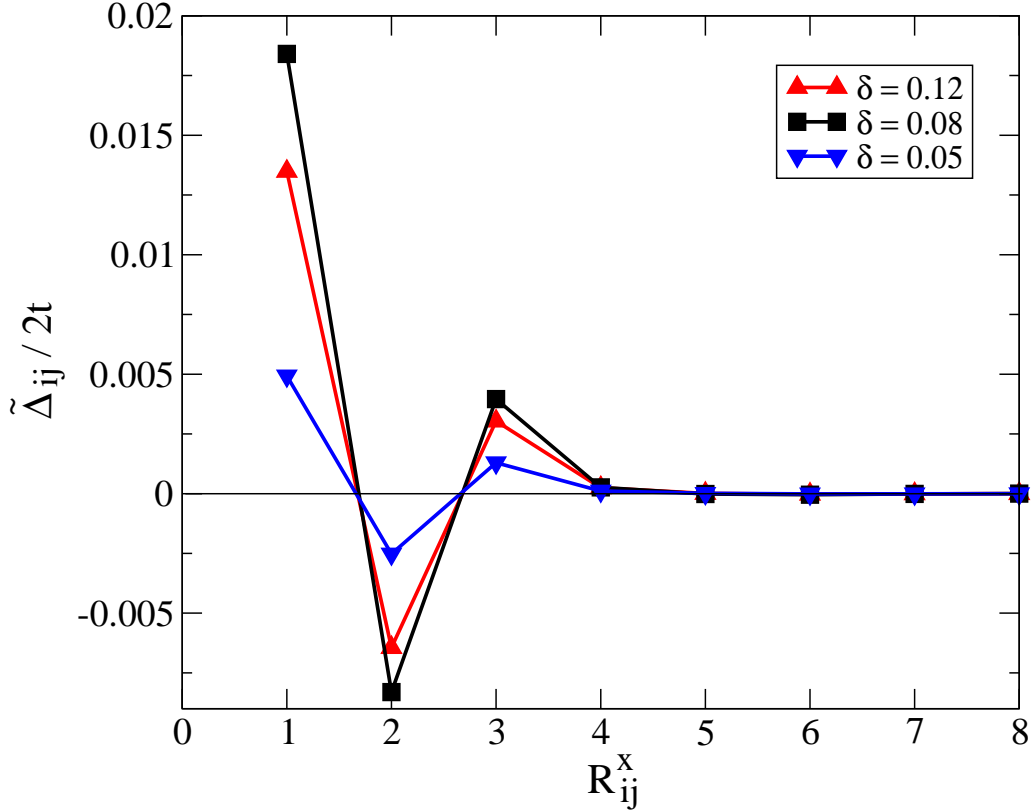}
    }
  \end{center}
  \caption{Superconducting order parameter $\tilde \Delta_{ij}$ (in units of $2t$) in local space 
along the $x$ direction (for $R_{ij}^y =0$) for three different hole fillings $\delta= 0.05$ (blue), $0.08$ (black), and $0.12$ (red).  The
parameters $t'$ and $T$ are the same as in Fig.~\ref{Fig_9}.
}
\label{Fig_13} 
\end{figure}

According to Fig.~\ref{Fig_9}, the gap function shows a pronounced ${\bf k}$-dependence 
in the whole Brillouin zone. By Fourier transforming  $\tilde\Delta_{\bf k}$ to the local space,
\begin{eqnarray}
 \label{71}
\tilde \Delta_{ij}&=& \frac{1}{N}\sum_{i,j} \tilde\Delta_{\bf k}^{(\infty)}\, 
 e^{i{\bf k}({\bf R}_i -{\bf R}_j)} \, ,
\end{eqnarray}
one finds the spatial dependence shown in Fig.~\ref{Fig_12}. The figure again 
reveals the $d$-wave character of the superconducting order parameter. Note that 
the strong ${\bf k}$-dependence of $\tilde{\Delta}_{\bf k}$ 
maps on a short range behavior in local space. As is clearly seen, the local
order parameter decays in space within a few lattice constants. 
This feature is consistent with the experimentally found superconducting coherence length in the cuprates 
of the order of a few lattice constants.  The order parameter changes its sign by proceeding along the $x$- or $y$-axis.
This can be seen for various hole fillings in Fig.~\ref{Fig_13}, where $\tilde \Delta_{ij}$ 
is shown as a function of $R_{ij}^x$ (for fixed $R_{ij}^y=0$).  Here  $R_{ij}^x$ and  $R_{ij}^y$ are 
the components of the difference vector
${\bf R}_{ij}= {\bf R}_i- {\bf R}_j$ between
lattice sites $ {\bf R}_i$ and $ {\bf R}_j$. 
The alternating sign of $\tilde{\Delta}_{ij}$
seems to be reminiscent of the sign behavior of antiferromagnetic correlations. However, the sign change
is a property of the superconducting state and not of 
antiferromagnetic correlations.

\subsubsection{Finite temperature results}

\begin{figure}
  \begin{center}
    \scalebox{0.55}{
      \includegraphics*{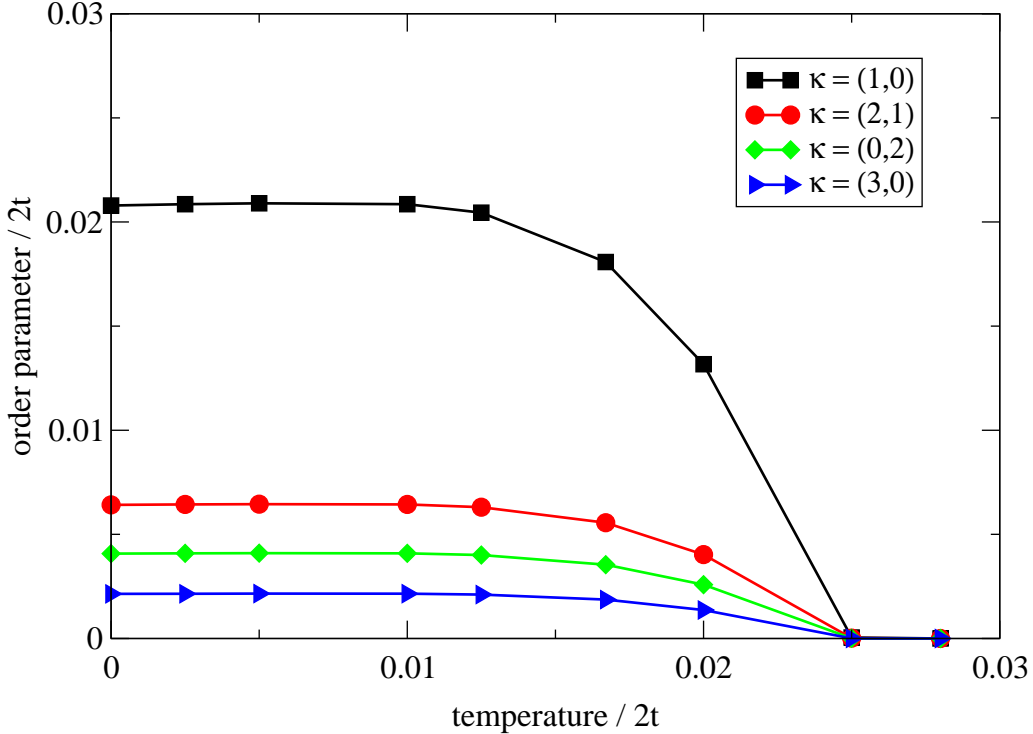}
    }
  \end{center}
  \caption{Local order parameter $\tilde \Delta_{ij} (T)$ as function of  
$T$ (both in units of $2t$) for different values of the distance 
$\kappa = |{\bf R}_{ij}|$. Note that all curves 
vanish at the same critical temperature $T_c$.
  }
  \label{Fig_14}
  \end{figure}

In Fig.~\ref{Fig_14}, the local order parameter $\tilde \Delta_{ij}$ is plotted 
as a function of $T$ for different values of the distance  
 between local sites, $\kappa = |{\bf R}_{ij}|$. The curves are  obtained from Fourier back transforming
Eq.~\eqref{71} together with the temperature dependent expression for 
$\langle \hat c_{-{\bf k}\downarrow}\hat c_{{\bf k}\uparrow} \rangle$ from Sec.~\ref{expvalues}. 
All curves vanish at the same temperature $T/2t \approx 0.026$, 
which defines the critical temperature $T_c$. 
Note that the temperature dependence  of  $\tilde \Delta_{ij}$ 
and thus of the gap function $\tilde \Delta_{\bf k}$ resembles that of the 
order parameter in BCS superconductors. This property can be traced back to the 
diagonalization approach on the basis of a Bogoliubov transformation in Appendix \ref{C}, 
which is applied to 
the renormalized Hamiltonian $\tilde{\cal H}$ in the superconducting state. 
Also the pair correlation function 
$\langle \hat c_{-{\bf k}\downarrow}\hat c_{{\bf k}\uparrow} \rangle$  is
evaluated in this way which results in a temperature dependence 
as in BCS superconductors as well.

\begin{figure}
 \begin{center}
    \scalebox{0.61}{
      \includegraphics*{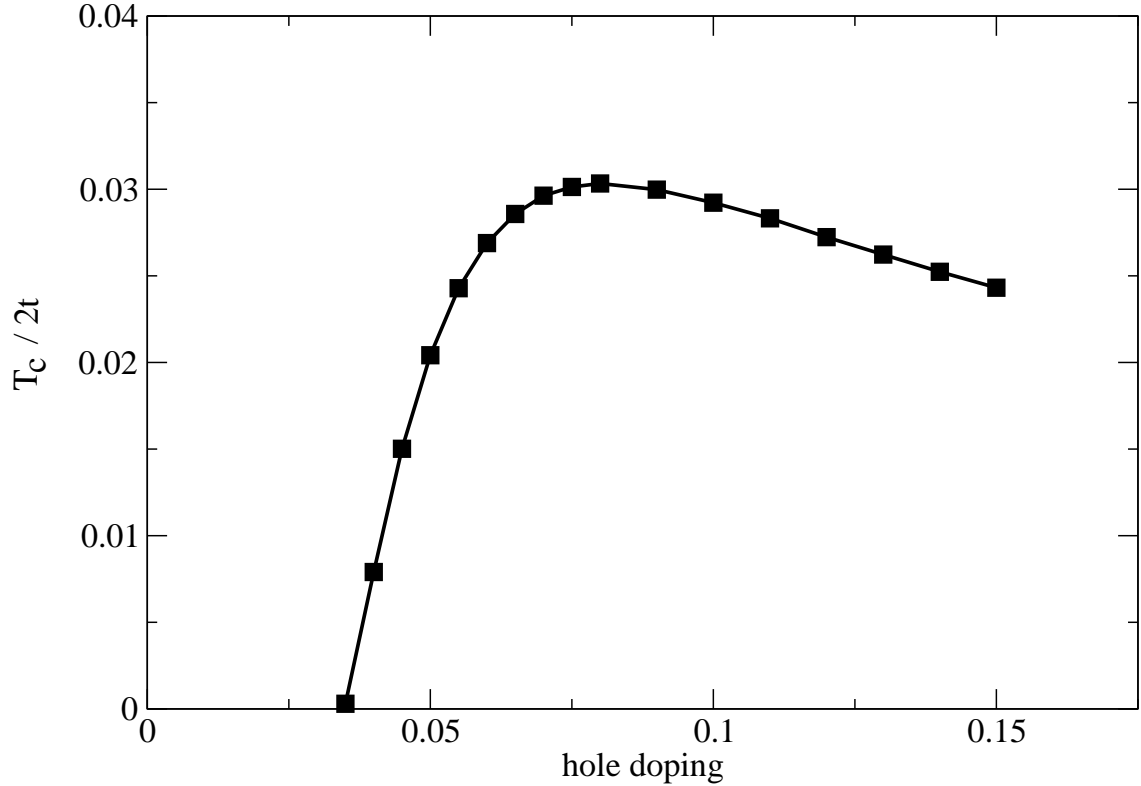}
    }
  \end{center}
  \caption{Critical temperature $T_c$ as a function of the hole doping
      $\delta$ for $t' = 0.4t$ and $J = 0.2t$. No
      superconducting solution is found for $\delta \le 0.03$. This result explains the vanishing of the 
superconducting phase in the cuprates at very low doping.}
\label{Fig_15} 
\end{figure}

In Fig.~\ref{Fig_15}, the critical temperature $T_c$ is given as a function of 
the hole doping $\delta$. The parameter values are again $t' = 0.4t$ and $J =
  0.2t$. Note that for small hole doping  $\delta \le 0.03$, 
no superconducting solutions are found. Also this result of  the PRM is in good agreement 
with experiments. In the underdoped region for $\delta > 0.03$, the critical temperature $T_c$ first
increases substantially until it arrives a maximum value at about  
$\delta \approx 0.08$. Above the optimal doping concentration of $\delta = 0.08$, 
the critical temperature decreases again (overdoped region).
Within the parameter range, given in the figure, the $T_c$ behavior agrees very well 
with experiment. For still larger values of $\delta$ ($\delta > 0.15$), our PRM result for   
$T_c$ remains finite. This feature is in disagreement with experiments, 
where the superconducting phase vanishes above a critical hole concentration. However,
this defect of the present approach is by no means surprising. As was discussed in 
Sec.~\ref{Re-app}, we have argued from the beginning that the present approach is not applicable
for the case of large hole doping. Nevertheless, Fig.~\ref{Fig_15} demonstrates that 
we are able to explain the experimental findings at least in the underdoped 
and in the optimal doping regime. For the present parameter values, the maximum of $T_c$  
at optimal doping is approximately given by $T_c \approx 0.06t$. 
Assuming a  bare bandwidth of $8t \approx 10^4 K$, this $T_c$-value corresponds to a critical 
temperature of order $50-100 K$, which is in the correct order of magnitude.

\subsubsection{Discussion}
Next, we want to discuss the origin of the superconducting pairing mechanism. Let us start with the 
superconducting order parameter $\tilde\Delta_{\bf k}^{(1)}$ after the 
first renormalization step. According to Eq.~\eqref{45}, we have
\begin{eqnarray}
\label{67}
\tilde {\Delta}_{{\bf k}}^{(n=1)} &=& {\Delta}_{{\bf k},\lambda=0} 
-\frac{1}{N} \sum_{\bf q} \frac{3J_{\bf q}}
{4 \hat \omega^2_{{\bf q},\lambda=0}} (\varepsilon_{{\bf k},\lambda=0} - 
\varepsilon_{{\bf k}+ {\bf q},\lambda=0} )^2\, 
\langle \hat c_{-({\bf k}+{\bf q})\downarrow}\, 
\hat c_{{\bf k}+{\bf q}\uparrow}\rangle .
\end{eqnarray}
The first term on the right hand side results from second order 
renormalization contributions 
according to Eq.~\eqref{36}. The numerical evaluation of Eq.~\eqref{67} shows that
it is small compared to the second term.  
According to Sec.~\ref{Re-app}, the latter one results from the factorization of the contribution 
$\sim \dot{\bf S}_{\bf q}\dot{\bf S}_{-{\bf q}}$
in the renormalized Hamiltonian ${\cal H}_{\lambda=0}=\sum_{\bf q} 
({J_{\bf q}})/({2 \hat \omega_{\bf q}^2}) 
\dot{\bf S}_{\bf q}\dot{\bf S}_{-{\bf q}} + \cdots $ 
after the first renormalization cycle.
Therefore, we can conclude from \eqref{A4} that the dominant part of the microscopic pairing 
interaction is given by 
\begin{eqnarray}
\label{69b} 
 {\cal H}_{(SC)} &=& \frac{1}{N}\sum_{{\bf q}{\bf k}} 
\frac{J_{\bf q}}{4\hat \omega_{\bf q}^2} (\varepsilon_{\bf k} - \varepsilon_{{\bf k}-{\bf q}})^2
\big( \hat c_{{\bf k} \uparrow}^\dagger \hat c_{-{\bf k} \downarrow}^\dagger 
\hat c_{-({\bf k}-{\bf q}) \downarrow} \hat c_{{\bf k}-{\bf q} \uparrow} 
+
2 \hat c_{{\bf k} \uparrow}^\dagger \hat c_{-{\bf k} \downarrow}^\dagger 
\hat c_{{\bf k}-{\bf q} \downarrow} \hat c_{-({\bf k}-{\bf q}) \uparrow} \big) \, .
\end{eqnarray}
Here, spin-singlet pairing was assumed. The expression \eqref{69b} is our central 
result for the superconducting pairing mechanism in the cuprates.
In contrast to usual BCS superconductors, where the pairing interaction 
between Cooper electrons is mediated by 
phonons, the present result can not be interpreted as an effective interaction of second order in some 
electron-bath coupling. Note that Eq.~\eqref{69b}
results from the part of the exchange ${\cal H}_J $ 
which commutes with ${\cal H}_t$.  
An important feature of the pairing 
interaction is the oscillation frequency 
$\hat{\omega}_{\bf q}^2$ in the denominator of Eq.~\eqref{69b},
 \begin{eqnarray}
\label{70}
\hat \omega^2_{\bf q} &=&
 - 2P_0(t^2_{{\bf q}=0}- t^2_{{\bf q}}) =  \hat \omega^2_{-{\bf q}} \geq 0 ,\quad \qquad
t^2_{\bf q} = 
\sum_{l (\neq i)}t_{il}^2\, \cos{{\bf q}({\bf R}_l - {\bf R}_i)} \, ,
\end{eqnarray}
which enhances the pairing mechanism for small hole doping, since $P_0 \sim \delta$.  
Therefore, the pairing interaction is mediated by oscillating hopping processes between nearest
neighbors. This was discussed in detail in Sec.~IV~A of paper I.  
First, an electron hops to a neighboring site which is empty. In the second step, it 
hops back to the first site, since this site was certainly empty after the 
first hop. Thereby, the presence of  short range antiferromagnetic correlations 
in the unperturbed Hamiltonian ${\cal H}_0$ is crucial, since it prevents the hopping 
to more distant sites. 

In order to derive an approximate gap equation, let us again start from Eq.~\eqref{67}.
When we restrict ourselves to a weak coupling theory, the $\lambda$-dependence of 
$\varepsilon_{{\bf k},\lambda}$ and $\hat{\omega}_{{\bf q},\lambda}$ can be neglected:
\begin{eqnarray}
\tilde {\Delta}_{{\bf k}}^{(1)} &=& 
-\frac{1}{N} \sum_{\bf q} \frac{3J_{\bf q}}
{4 \hat \omega^2_{\bf q}} (\varepsilon_{\bf k} - 
\varepsilon_{{\bf k}+ {\bf q}} )^2\, 
\langle \hat c_{-({\bf k}+{\bf q})\downarrow}\, 
\hat c_{{\bf k}+{\bf q}\uparrow}\, 
\rangle  \, ,
\label{70a}
\end{eqnarray}
where the first term from Eq.~\eqref{67} was already omitted. 
For a purely qualitative discussion of the gap parameter, 
let us abandon all higher order renormalization effects, 
which would be included in the full renormalization 
scheme of Sec.~\ref{Re-app}.
Inserting the former expression \eqref{64} for 
$\langle \hat c_{-{\bf k}\downarrow} \hat c_{{\bf k} \uparrow}\rangle$
into Eq.~\eqref{70a}, we find 
\begin{eqnarray}
\label{68}
\tilde {\Delta}_{{\bf k}}^{(1)} &=&
-\frac{1}{N} 
 \sum_{\bf q} \frac{3J_{\bf q}}
{4 \hat \omega^2_{\bf q}} (\varepsilon_{\bf k} - 
\varepsilon_{{\bf k}+ {\bf q}} )^2\,  
\tilde u_{{\bf k}+{\bf q}}^2 D^2
\frac{1 -2f(E_{{\bf k}+{\bf q}})}{2 \sqrt{\varepsilon_{{\bf k}+{\bf q}}^{2} 
+ D^2 \tilde \Delta_{{\bf k}+{\bf q}}^2 } } 
\tilde \Delta _{{\bf k}+{\bf q}}  ,
\end{eqnarray}
where $E_{\bf k}$ is again given by 
$E_{\bf k} = \sqrt{\varepsilon_{\bf k}^{2} + D^2
\tilde \Delta_{\bf k}^2 }$, and $f(E_{\bf k})$ is the Fermi function 
$f(E_{\bf k}) = 1 / (1 + e^{\beta E_{\bf
  k}})$.
Moreover, by replacing on the left hand side also $\tilde{\Delta}_{\bf k}^{(1)}$
by $\tilde{\Delta}_{\bf k}$, we arrive at the following approximate gap equation  
\begin{eqnarray}
\label{69}
\tilde {\Delta}_{{\bf k}} &\approx& 
-\frac{1}{N} D^2
 \sum_{\bf q} \frac{3J_{\bf q}}
{4 \hat \omega^2_{{\bf q}}} (\varepsilon_{{\bf k}} - 
\varepsilon_{{\bf k}+ {\bf q}} )^2\,  
\tilde u_{{\bf k}+{\bf q}}^2 
\frac{1 -2f( E_{{\bf k}+{\bf q}})}{2 \sqrt{\varepsilon_{{\bf k}+{\bf q}}^{2} 
+ D^2 \tilde \Delta_{{\bf k}+{\bf q}}^2 } } 
\tilde \Delta _{{\bf k}+{\bf q}}.            
\end{eqnarray}
Note that the main features of our numerical results for the full renormalization scheme
 can already be detected from  this equation.  
Due to the doping dependence of $\tilde{u}_{\bf k}$, shown in Fig.~9 of paper I, 
superconductivity sets in at the same small $\delta$-value,
at which $\tilde u_{\bf k}$ becomes non-zero. 
With increasing hole doping, $\tilde u_{\bf k}$ increases, which also leads to a 
strengthening of the coherent excitation in $\Im G({\bf k},\omega)$.
Moreover, superconductivity is favored for low doping due to the factor 
$\hat \omega^2_{\bf q} \sim \delta$ in the denominator of Eq.~\eqref{69}. Both features together,
i.e.~the increase of $\tilde u_{\bf k}$ with $\delta$ and  $\hat \omega^2_{\bf q} \sim \delta$
lead to a maximum of $T_c$  at a finite doping value which is seen in 
Fig.~\ref{Fig_15}.
The property $\hat \omega^2_{\bf q} \sim \delta$
also explains the decrease of $T_c$ in the overdoped region, since renormalization processes
become weaker for larger $\delta$.

The preference of the PRM to find solutions with $d$-wave symmetry 
for the gap parameter
can also be understood from the gap equation \eqref{69}. For an explanation, 
let us start by dividing  
the sum over $\bf q$ in Eq.~\eqref{69}
into two parts with $|\varepsilon_{{\bf k}+{\bf q}}| \leq 
|\tilde \Delta_{{\bf k}+{\bf q}}|$
and $|\varepsilon_{{\bf k}+{\bf q}}| > |\tilde \Delta_{{\bf k}+{\bf q}}|$. 
Omitting the second sum, one finds  
\begin{eqnarray}
\label{70d}
\tilde {\Delta}_{{\bf k}} &\approx& 
-\frac{1}{N}\, \sum_{{\bf q}, 
|\varepsilon_{{\bf k}+{\bf q}}| \leq |\tilde \Delta_{{\bf k}+{\bf q}}|}
\frac{3J_{\bf q}} {4 \hat \omega^2_{\bf q}}\,
( \varepsilon_{\bf k} -  \varepsilon_{{\bf k}+{\bf q}})^2\, 
\tilde u_{{\bf k}+{\bf q}}^2 D^2
\frac{1 -2f( E_{{\bf k}+{\bf q}})}{2 \sqrt{\varepsilon_{{\bf k}+{\bf q}}^{2} 
+ \tilde \Delta_{{\bf k}+{\bf q}}^2 } } \,
\tilde \Delta _{{\bf k}+{\bf q}}  .          
\end{eqnarray}
For most values of ${\bf k}$, the neglected sum is smaller 
by a factor of order $\Delta/t$. 
An exception are ${\bf k}$-values close to the Fermi surface 
${\bf k}\approx {\bf k}_F$ (with $|\varepsilon_{\bf k}| \leq O(\Delta_{\bf k})$),
which will be excluded in the following discussion. Here, the
sum with $|\varepsilon_{{\bf k}+{\bf q}}| > |\tilde \Delta_{{\bf k}+{\bf q}}|$ 
would be larger by a factor of order $t/\Delta$. 
With respect to Eq.~\eqref{70d}, those terms of the ${\bf q}$ sum are most important, 
which have energies $|\varepsilon_{{\bf k}+{\bf q}}|$ not exceeding 
$|\tilde \Delta_{{\bf k}+{\bf  q}}|$. For ${\bf k}$-values
on the diagonal,  $k_x= k_y$,
of the Brillouin zone, it can be seen that ${\bf q}$-values with $q_y\approx q_x \pm \pi$ lead to
 small energies $\varepsilon_{{\bf k}+ {\bf q}}\approx 0$ and thus to the dominant 
contributions in Eq.~\eqref{70d}. Here, the dispersion relation
$\varepsilon_{\bf k}= -2t (\cos{k_x a} + 
\cos{k_y a})$ was used. However, the prefactor $J_{\bf q}$ vanishes in this case. 
This explains the nodal line $k_x= k_y$ and similarly  $k_x= -k_y$ of the gap parameter in 
Fig.~\ref{Fig_10}.
However note that the exchange constant $J_{\bf q}$ changes its sign as a
function of $\bf q$. From this behavior, one can
conclude that $d$-wave symmetry for the order parameter 
is more favorable than $s$-wave symmetry.


\subsection{ARPES Spectral functions}

Finally, let us discuss the ARPES spectral function in the
  superconducting phase. This quantity is obtained from the
  dissipative part of the anticommutator Green function
  \eqref{49}. 

\begin{figure}
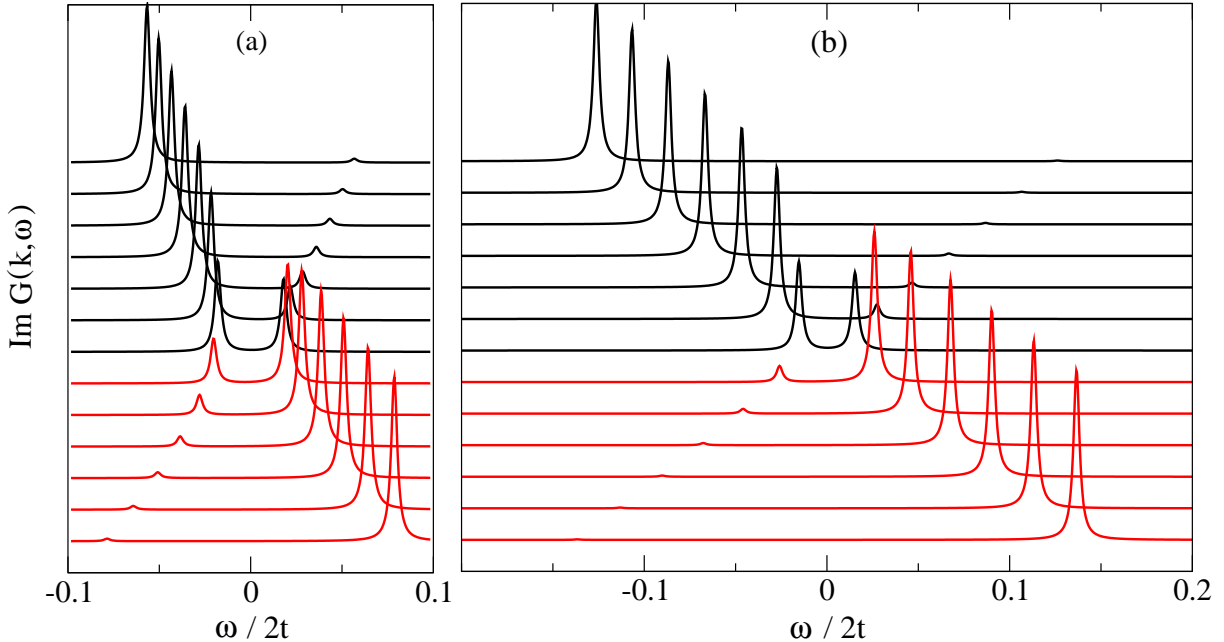

  \begin{center}
    \scalebox{0.5}{
      \includegraphics*{spektr_k_n92T0_a.eps}
      \includegraphics*{spektr_k_n92T0_m.eps}
    }
  \end{center}
  \caption{Spectral functions $\Im G({\bf k},\omega)$ in the superconducting
      phase at optimal doping, $\delta=0.08$
for two fixed $k_x$-values: (a) $k_x= \pi$ (anti-nodal region) and (b) $k_x= 5\pi/8$ (in between anti-nodal and
nodal region). By varying $k_y$ , the Fermi surface is crossed. The other parameters are $t' = 0.4t$, $J = 0.2t$,
and $T=0$.
}
  \label{Fig_17}
  \end{figure}

\begin{figure}
  \begin{center}
    \scalebox{0.5}{
      \includegraphics*{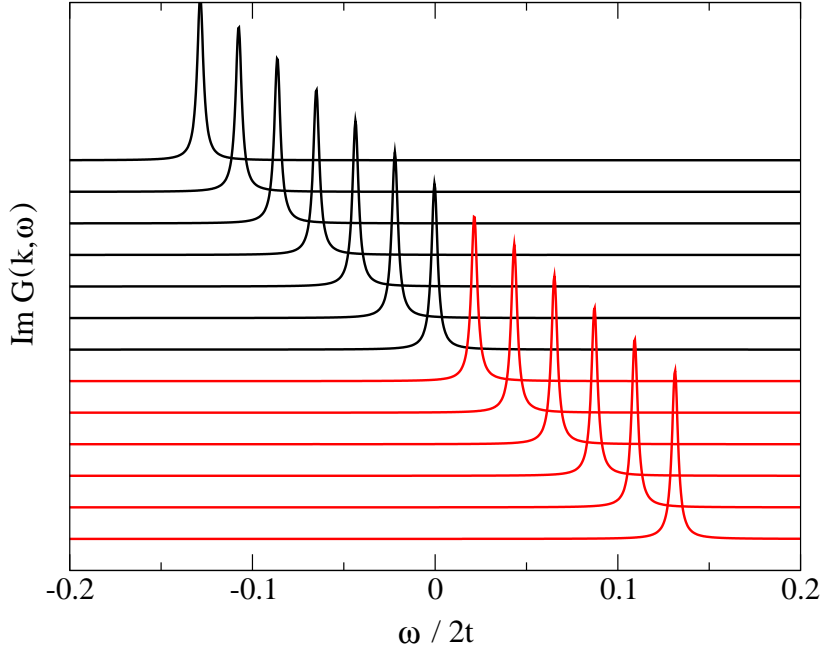}
    }
  \end{center}
  \caption{Spectral functions $\Im G({\bf k},\omega)$ as in Fig.~\ref{Fig_17}
  for a fixed $k_x$-value, $k_x= \pi/2$ . By varying $k_y$ the Fermi surface is crossed
in the nodal region.
}
  \label{Fig_18}
  \end{figure}

\begin{figure}
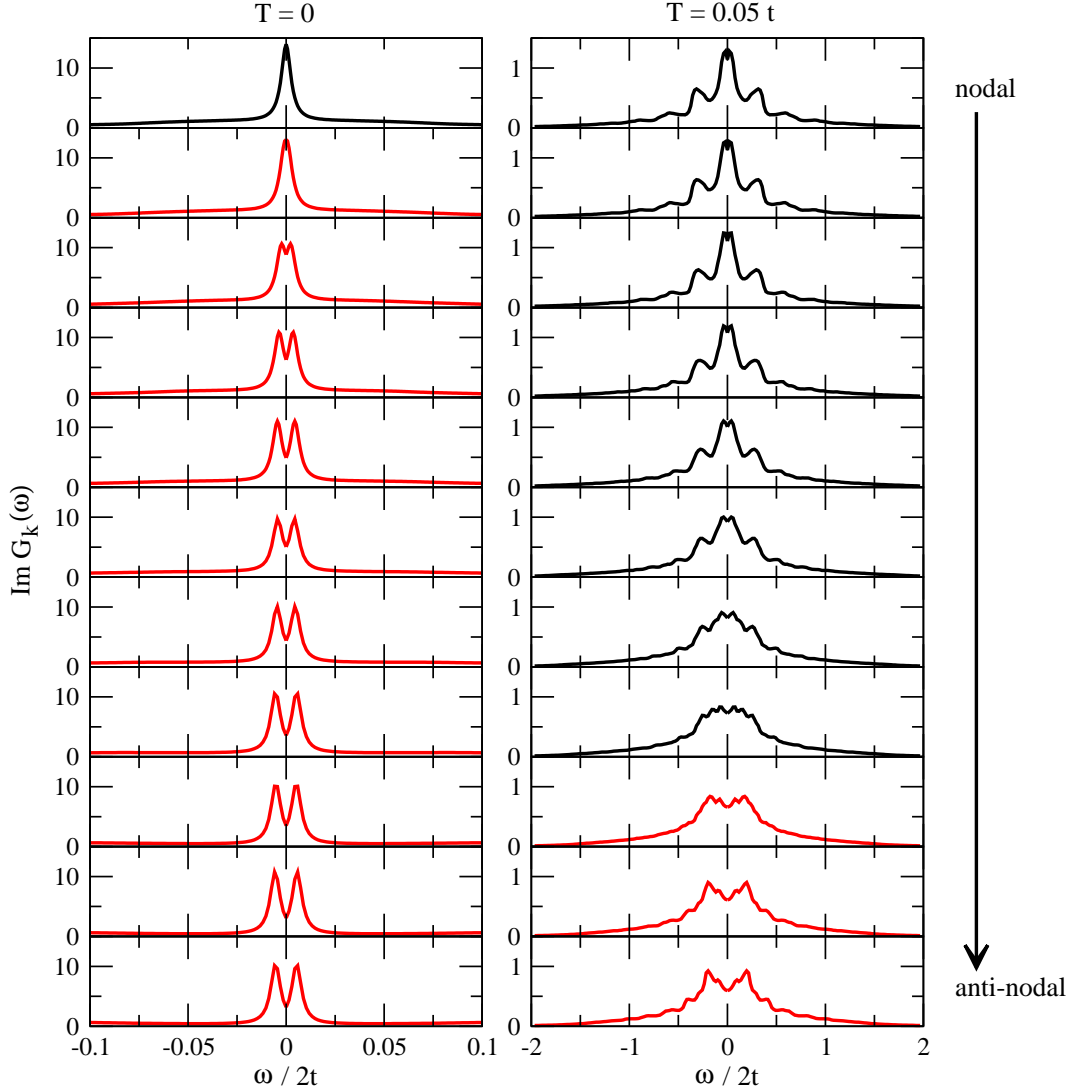

  \begin{center}
    \scalebox{0.7}{
      \includegraphics*{spektr_fl_n95T0.eps}
      \includegraphics*{spektr_fl_n95T025.eps}
    }
  \end{center}
  \caption{
   Symmetrized spectral functions $\Im G({\bf k},\omega)$ for ${\bf k}$-values on the FS 
between the nodal (top) and anti-nodal point (bottom) for two temperatures
(a) in the superconducting phase at $T=0$, and (b) in the pseudogap phase 
at $T=0.05t$. The critical temperature is $T_c = 0.03t$ (underdoped case $\delta = 0.05$). The
other parameters are the same as in Fig.~\ref{Fig_17}. 
  }
  \label{Fig_19}
  \end{figure}

In Figs.~\ref{Fig_17} - \ref{Fig_19}, our results for the superconducting phase
are given which are obtained from the numerical evaluation of
Eq.~\eqref{88}. First, in
Fig.~\ref{Fig_17}, we have chosen as parameters: $\delta=0.08$
(optimal doping), $T=0$, $t' = 0.4t$, and $J = 0.2t$. Two cuts 
with fixed $k_x$ and varying $k_y$ are shown. Thereby the FS is crossed. 
In panel (a), where $k_x=\pi$, the spectra belong to ${\bf k}$-values  in 
the anti-nodal region, whereas in (b) $k_x= 5\pi/8$. Here, a ${\bf k}$-region is probed 
in-between the nodal and the anti-nodal point.  The 
spectra in both panels display peak-like structures in a small
  energy range around $\omega=0$. Note that all structures are caused alone  
  by the coherent part of  $\Im G({\bf k},\omega)$ (first line in Eq.~\eqref{88}),
which  consists of two peaks at the positions $\omega = \pm
  E_{\bf k}$. For ${\bf k}$-vectors, far away from the FS (top and bottom plots 
in Figs.~\ref{Fig_17}(a) and (b)),
  a dominating peak at $\omega \approx \tilde\varepsilon_{\bf k}$ is found, which arises from 
the excitations at $ \pm E_{\bf k}$, depending on the sign of $\tilde\varepsilon_{\bf k}$. 
 By approaching the FS, a secondary peak arises at $\omega \approx
  -\tilde \varepsilon_{\bf k}$. 
  An expansion of the prefactors in Eq.~\eqref{88} shows that in each case the secondary 
  peak has a smaller weight of order 
  $(\tilde \Delta_{\bf k} / \tilde \varepsilon_{\bf k})^2$. 
  Only for ${\bf k}$-values on the FS ($\tilde \varepsilon_{\bf k} = 0$), 
  the two coherent peaks have equal weight. 
  They are separated by an energy distance, which is given by the gap parameter 
  $(2D\tilde \Delta_{\bf k})$. Note that the gap size is almost the same for the two  cases of
  Fig.~\ref{Fig_17}.
   A comparison of both panels of Fig.~\ref{Fig_17} also shows that the secondary
  peak is more pronounced in the anti-nodal region than in-between 
the anti-nodal and nodal region.
   Furthermore, the overall dispersion of $\tilde
  \varepsilon_{\bf k}$  of the primary
  peak is weaker in the anti-nodal region than for the case of intermediate $k_x$-values.
  With respect to the incoherent contributions to  $\Im G({\bf k},\omega)$, 
  note that for optimal doping the overall weight of the
  coherent and of the incoherent excitations are approximately the same. 
However, the incoherent part of the spectrum is spread over a 
much larger frequency range. Therefore, in a small $\omega$-range, close to the Fermi level, 
the coherent excitations are dominant.

In Fig.~\ref{Fig_18}, the spectral function is
  plotted in the nodal region  for fixed $k_x= \pi/2$  
and different values of $k_y$. Thereby, again the FS is crossed.  
Note that neither a secondary peak nor a
  superconducting gap is found in the nodal region. Also,  the coherent peak moves
  almost unchanged through the FS, when $k_y$ is varied.

Finally, Fig.~\ref{Fig_19} shows the results for the symmetrized spectral functions $\Im G({\bf k},\omega)$
for two different temperatures (a) $T = 0$ (superconducting
phase)  and (b) $T = 0.05t$ (pseudogap phase). The  $\bf k$-values proceed
on the FS between the nodal (top) and the anti-nodal (bottom)
point. The hole concentration is $\delta = 0.05$ (underdoped regime) which leads to 
a critical temperature $T_c = 0.03t$. 
In the spectra at  temperature $T=0$, one recognizes the opening of a 
superconducting gap for all $\bf k$-vectors
except at the nodal point. The gap size as a function of the Fermi
surface angle $\phi$ is given by the blue line 
in Fig.~\ref{Fig_11}.
Similar as before, the peak-like structure  
arises from the coherent excitations in a
small $\omega$-range around $\omega=0$. For the higher temperature, $T=0.05t$ (pseudogap phase),
the system is in the normal state. On a substantial part of the Fermi surface,
the spectra now show the typical large spectral weight around $\omega = 0$,
indicating a Fermi arc of gapless excitations. The Fermi arc extents 
over a finite ${\bf k}$-range. In contrast to the superconducting case, 
the spectrum is now
dominated by the incoherent excitations. In the anti-nodal region, they form the 
pseudogap around $\omega = 0$ (see also paper I). 
Note that the pseudogap in Fig.~\ref{Fig_19}(b) is about ten
times larger than the superconducting gap at $T = 0$ (for the present hole doping  
$\delta=0.05$). Note that 
for both temperatures,  the spectra are in good qualitative agreement with
recent  ARPES measurements\cite{K06},\cite{K07},\cite{K08}.

Let us finally make one remark concerning the linewidth of the coherent
peaks. As was already mentioned in Sec.~V of paper I, from the experimental
point of view, we would expect a temperature
dependent broadening of the coherent peaks which is caused by the coupling to other degrees of
freedom. Such a broadening was not incorporated in the present approach. 
Note, however, that a broadening of the spectra is also produced by 
the incoherent excitations of $\Im G({\bf k},\omega)$. In order to include a 
temperature dependent broadening of the coherent excitations, 
we have added by hand a small linewidth in Fig.~\ref{Fig_19}, which is taken  
of the order of $k_B T$.


\section{Conclusions}
In this paper, we have given a microscopic approach to the
superconducting phase in cuprate systems at moderate hole doping. Thereby, a recently 
developed projector-based renormalization method (PRM) was applied to the
$t$-$J$ model.
Our result for the superconducting order parameter shows $d$-wave 
symmetry with a coherence length of a few lattice constants which is in
agreement with experiments. In contrast to usual BCS superconductors, where the pairing interaction 
between Cooper electrons is mediated by 
phonons, the superconducting pairing interaction in the cuprates 
can not be interpreted as an effective interaction of second order in some 
electron-bath coupling. Instead, 
the main contribution to the pairing results from 
the part of the exchange interaction which commutes with the hopping Hamiltonian ${\cal H}_t$. 
The superconducting state naturally arises from a
typical oscillation behavior of the correlated electrons between neighboring
lattice sites due to the presence of spin fluctuations. 
The theoretical results can explain the experimental findings in the underdoped 
as well as in the optimal doping regime. The obtained value of $T_c$  at optimal doping 
has the correct order of magnitude.

\section{Acknowledgments}
We would like to acknowledge stimulating and 
enlightening discussions with J.~Fink and
A.~H\"ubsch. This work was supported by 
the DFG through the research program SFB 463.


 \begin{appendix}

\section{Factorization approximation for  $\dot{\bf S}_{\bf q}\dot{\bf S}_{-{\bf q}}$}
\label{A}
The aim of this appendix is to simplify the operator product 
$\dot{\bf S}_{\bf q}\dot{\bf S}_{-{\bf q}}$ which enters the expressions \eqref{29a} for 
${\cal H}_{0,\lambda}$ and ${\cal H}_{1,\lambda}$. 
As in paper I, we start from the expression 
\begin{eqnarray}
\label{A1}
\dot{\bf S}_{{\bf q}}\dot{\bf S}_{-{\bf q}}
=  \frac{1}{4N}  
\sum_{\alpha \beta} \sum_{\gamma \delta}
(\vec \sigma_{\alpha \beta}\cdot \vec \sigma_{\delta \gamma}) 
\sum_{i \neq j}t_{ij}(e^{i {\bf q}{\bf R}_i} - e^{i {\bf q}{\bf R}_j})
\sum_{l \neq m} t_{lm}(e^{-i {\bf q}{\bf R}_l} - e^{-i {\bf q}{\bf R}_m})\,
\hat c_{i\alpha}^\dagger \hat c_{j\beta}\hat c_{m\delta}^\dagger \hat c_{l\gamma}.  \nonumber \\
&&
\end{eqnarray}
The four-fermion operator on the right hand side 
can be factorized in two different ways: One can either reduce it to operators  
 $\hat c^\dagger_{{\bf k}\sigma} \hat c_{{\bf k}\sigma}$ or to operators
$\hat c^\dagger_{{\bf k}\sigma} \hat c^\dagger_{-{\bf k}, -\sigma}$ and
$\hat c_{-{\bf k}, -\sigma} \, \hat c_{{\bf k}, \sigma}$.  The first factorization 
will lead to a renormalization of $\varepsilon_{\bf k}$, whereas the second one 
renormalizes the superconducting order parameter $\Delta_{\bf k}$.
In the factorization, we have to pay attention to the fact that the averaged spin operator 
vanishes $\langle {\bf S}_i \rangle =0$ outside the antiferromagnetic regime. 
Moreover, all local indices in the four-fermion term of \eqref{A1} should 
be different from each other. This follows from the former 
decomposition of the exchange interaction into eigenmodes of ${\sf L}_t$, 
where we have implicitly assumed that the 
operators $\dot {\bf S}_{{\bf q}}$ and $\dot {\bf S}_{-{\bf q}}$ 
do not overlap in the local space. Otherwise, the decomposition would be much more involved.
However, it can be shown that these 'interference' terms only make a minor impact on the 
results. 
 
\noindent
(i) For the 'normal' factorization, we find
\begin{eqnarray}
\label{A3}
 && 
\dot{\bf S}_{{\bf q}}\dot{\bf S}_{-{\bf q}}|_{(\mbox {\tiny i})} 
= 
 -\frac{3}{2N} \sum_{{\bf k}\sigma}
( \varepsilon_{\bf k} -  \varepsilon_{{\bf k}-{\bf q}} )^2
\langle (\hat c_{{\bf k}-{\bf q}\alpha}^\dagger \hat c_{{\bf k}-{\bf q}\alpha})_{NL} \rangle \,
(\hat c_{{\bf k}\sigma}^\dagger \hat c_{{\bf k}\sigma})_{NL}  \, ,
\end{eqnarray}
where we have defined 
$ (\hat c_{{\bf k}\sigma}^\dagger \hat c_{{\bf k}\sigma})_{NL}  =
\hat c_{{\bf k}\sigma}^\dagger \hat c_{{\bf k}\sigma}  
- ({1}/{N}) \sum_{{\bf k}'}\hat c_{{\bf k}'\sigma}^\dagger \hat c_{{\bf k}'\sigma}$.
The attached subscript ${NL}$  
indicates that the local sites of the operators 
inside the brackets are different from each other. 
In Eq.~\eqref{A3}, we have also neglected an additional 
c-number quantity, which enters in the factorization, and the  
sums over the spin indices in Eq.~\eqref{A1} have already been 
carried out 

\noindent
(ii) By assuming spin-singlet pairing, we obtain from Eq.~\eqref{A1}, 
\begin{eqnarray}
\label{A4} 
\dot{\bf S}_{{\bf q}}\dot{\bf S}_{-{\bf q}}|_{(\mbox{\tiny {ii}})} 
 &=& \frac{1}{2N}\sum_{{\bf k}} 
 (\varepsilon_{\bf k} - \varepsilon_{{\bf k}-{\bf q}})^2
\big( \hat c_{{\bf k} \uparrow}^\dagger \hat c_{-{\bf k} \downarrow}^\dagger 
\hat c_{-({\bf k}-{\bf q}) \downarrow} \hat c_{{\bf k}-{\bf q} \uparrow} 
+
2 \hat c_{{\bf k} \uparrow}^\dagger \hat c_{-{\bf k} \downarrow}^\dagger 
\hat c_{{\bf k}-{\bf q} \downarrow} \hat c_{-({\bf k}-{\bf q}) \uparrow} \big). 
\end{eqnarray} 
According to Sec.~IV~A, the expression \eqref{A4} leads 
to the main part of the superconducting pair interaction. In a factorization approximation, 
the two contributions in \eqref{A4} can be combined to
\begin{eqnarray}
\label{A5}
  && \dot{\bf S}_{{\bf q}}\dot{\bf S}_{-{\bf q}}|_{(\mbox{\tiny {ii}})} 
= 
 \frac{3}{2N}  \sum_{{\bf k}}
( \varepsilon_{\bf k} -  \varepsilon_{{\bf k}-{\bf q}} )^2
\left\{
\langle \hat c_{-({\bf k}-{\bf q})\downarrow } \hat c_{{\bf k}-{\bf q}\uparrow} \rangle \, \,
\hat c_{{\bf k}\uparrow}^\dagger \hat c_{-{\bf k}\downarrow}^\dagger 
+ h.c. \right\}.
\end{eqnarray}
Using Eqs.~\eqref{A3} and \eqref{A5} together with Eq.~\eqref{43}, one is finally led 
to the renormalization result \eqref{45}   
for $\tilde{\varepsilon}_{\bf k}^{(0)}$ and 
$\tilde{\Delta}_{\bf k}^{(0)}$ to first order in $J$.

The above factorization can also be used to derive  the 
renormalization contributions \eqref{35},\eqref{36}
to ${\varepsilon}_{{\bf k},\lambda -\Delta \lambda}$  
and $\Delta_{{\bf k},\lambda - \Delta \lambda}$. Using the expressions 
\eqref{A3} and \eqref{A4}, we can first simplify the second order renormalization
${\cal H}_{\lambda- \Delta\lambda}^{(2)}$ of ${\cal H}_{\lambda - \Delta \lambda}$.
In analogy to the results of Appendix B in paper I,  we arrive at  
\begin{eqnarray}
\label{A9}
 {\cal H}_{\lambda - \Delta \lambda}^{(2)} &=& 
3\sum_{\bf q}
(\frac{J_{\bf q}}{4 \hat \omega_{\bf q}^2})^2\, \Theta_{\bf q}(\lambda, \Delta \lambda)
\left( 
\left[
\frac{1}{N} \sum_{{\bf k}\sigma}(2\varepsilon_{\bf k} -
\varepsilon_{{\bf k}+{\bf q}} -\varepsilon_{{\bf k}-{\bf q}}) 
\langle \hat c_{{\bf k}\sigma}^\dagger  \hat c_{{\bf k}\sigma} \rangle 
\right]\,  {\bf S}_{\bf q}\cdot {\bf S}_{-{\bf q}} \right. \nonumber  \\
&& \left. \qquad +
 \langle {\bf S}_{\bf q}\cdot {\bf S}_{-{\bf q}} \rangle \,
 \frac{1}{N} \sum_{{\bf k}\sigma}(2\varepsilon_{\bf k} -
\varepsilon_{{\bf k}+{\bf q}} -\varepsilon_{{\bf k}-{\bf q}}) \;
 \hat c_{{\bf k}\sigma}^\dagger  \hat c_{{\bf k}\sigma} \right)
 \nonumber  \\
&-& \sum_{\bf q}
 (\frac{J_{\bf q}}{4 \hat \omega_{\bf q}^2})^2\, \Theta_{\bf q}(\lambda, \Delta \lambda)
 \langle \dot {\bf S}_{\bf q}\cdot \dot {\bf S}_{-{\bf q}} \rangle \,
 \frac{1}{N} \sum_{{\bf k}\sigma}(2\varepsilon_{\bf k} -
\varepsilon_{{\bf k}+{\bf q}} -\varepsilon_{{\bf k}-{\bf q}}) \;
 \hat c_{{\bf k}\sigma}^\dagger  \hat c_{{\bf k}\sigma} 
 \nonumber  \\
&+&
\frac{3}{2N} \sum_{{\bf q}\sigma} (\frac{J_{\bf q}}{4 \hat \omega_{\bf q}^2})^2 \,
\Theta_{\bf q}(\lambda,\Delta \lambda)\,
\left[ \frac{1}{N} \sum_{{\bf k}'\sigma'}(2\varepsilon_{{\bf k}'} -
\varepsilon_{{\bf k}'+{\bf q}} -\varepsilon_{{\bf k}'-{\bf q}}) 
\langle \hat c_{{\bf k}'\sigma'}^\dagger  \hat c_{{\bf k}'\sigma'} \rangle 
\right] \times \nonumber \\
&& \qquad \times \sum_{{\bf k}\sigma}   (\varepsilon_{\bf k}- \varepsilon_{{\bf k}- {\bf q} })^2
\langle (\hat c_{{\bf k}-{\bf q}\alpha}^\dagger \hat c_{{\bf k}-{\bf q}\alpha})_{NL} \rangle
\, (\hat c_{{\bf k}\sigma}^\dagger \hat c_{{\bf k}\sigma} )_{NL}  \nonumber \\
&-& \frac{1}{2N}
\sum_{{\bf q}\sigma} (\frac{J_{\bf q}}{4 \hat \omega_{\bf q}^2})^2 \,
\Theta_{\bf q}(\lambda,\Delta \lambda) \,
\left[ \frac{1}{N} \sum_{{\bf k'}\sigma'}(2\varepsilon_{{\bf k}'} -
\varepsilon_{{\bf k}'+{\bf q}} -\varepsilon_{{\bf k}'-{\bf q}}) 
\langle \hat c_{{\bf k}'\sigma'}^\dagger  \hat c_{{\bf k}'\sigma'} \rangle 
\right] \times \nonumber \\
&&  \qquad \times  \sum_{\bf k} (\varepsilon_{\bf k}- \varepsilon_{{\bf k}- {\bf q} })^2
\langle (\hat c_{{\bf k}-{\bf q}\downarrow} \hat c_{{\bf k}-{\bf q}\uparrow}) \rangle
\, (\hat c_{{\bf k}\uparrow}^\dagger \hat c_{{\bf k}\downarrow}^\dagger + h.c. )  .
\end{eqnarray}
From \eqref{A9}, the second order renormalizations to
$\varepsilon_{{\bf k},\lambda - \Delta \lambda}$ and $\Delta_{{\bf k},\lambda - \Delta \lambda}$ 
can immediately be deduced.

\section{Bogoliubov transformation for the superconducting Hamiltonian $\tilde{\cal H}$}
\label{C}
The aim of this appendix is to diagonalize the renormalized Hamiltonian $\tilde {\cal H}$
for the superconducting phase. According to Eq.~\eqref{46}, the Hamiltonian $\tilde {\cal H}$ reads 
\begin{eqnarray} 
 \label{C1}
\tilde {\cal H} &=& \sum_{{\bf k}\sigma} \tilde \varepsilon_{\bf k}\,
\hat c_{{\bf k} \sigma}^\dagger \, \hat c_{{\bf k} \sigma}
- \sum_{\bf k} \left( \tilde {\Delta}_{{\bf k}}\, \hat c_{{\bf k}, \uparrow}^\dagger 
\hat c_{-{\bf k}, \downarrow}^\dagger + \tilde {\Delta}_{{\bf k}}^{*}\, 
\hat c_{-{\bf k}, \downarrow} \hat c_{{\bf k}, \uparrow}
\right)     + \tilde E        .
\end{eqnarray}
Due to the presence of the Hubbard operators in Eq.~\eqref{C1}, 
the usual Bogoliubov transformation can only be applied 
approximately. Let us start by introducing new fermion operators,
\begin{eqnarray}
\label{C2}
\alpha_{\bf k}^\dag &=& {\sf U}_{\bf k}\, \hat c_{{\bf k},\uparrow}^\dag - {\sf V}_{\bf k}\,
\hat c_{-{\bf k},\downarrow},  \\
\beta_{\bf k}^\dag &=& {\sf U}_{\bf k}\, \hat c_{-{\bf k},\downarrow}^\dag + {\sf V}_{\bf k}\,
\hat c_{{\bf k},\uparrow}. \nonumber  
\end{eqnarray}
We require that $\alpha_{\bf k}^\dag$ and $\beta_{\bf k}^\dag$ are eigenmodes of
$\tilde{\cal H}$,
\begin{eqnarray}
\label{C3}
\tilde{\sf L} \alpha_{\bf k}^\dag &=& E_{\bf k} \alpha_{\bf
  k}^\dag \, , \quad
\tilde{\sf L} \beta_{\bf k}^\dag = E_{\bf k} \beta_{\bf  k}^\dag. 
\end{eqnarray}
In order to find equations for ${\sf U}_{\bf k}$ and ${\sf V}_{\bf k}$, 
let us insert the expression \eqref{C2} for 
$\alpha_{\bf k}^\dagger$ into the first equation of \eqref{C3},
\begin{eqnarray}
\label{C4}
{\sf U}_{\bf k} \, \tilde{\sf L} \hat c_{{\bf k},\uparrow}^\dag - 
{\sf V}_{\bf  k} \, \tilde{\sf L} \hat c_{-{\bf k},\downarrow} &=& E_{\bf k}
  \left( {\sf U}_{\bf k}\, \hat c_{{\bf k},\uparrow}^\dag - {\sf V}_{\bf k}\,
\hat c_{-{\bf k},\downarrow} \right) .
\end{eqnarray}
The two commutators on the left hand side of Eq.~\eqref{C4} will be evaluated separately.    
For the first one, $\tilde{\sf L} \hat c_{{\bf k}\sigma} = [\tilde{\cal H},\hat c_{{\bf
      k},\uparrow}^\dag]$, we obtain:
\begin{eqnarray*}
\tilde{\sf L} \hat c_{{\bf k},\uparrow}^\dag &=& 
\tilde{\sf L}_t \hat c_{{\bf k},\uparrow}^\dag 
- \sum_{{\bf k}'}  \tilde {\Delta}_{{\bf k}'}^*
      [ \hat c_{-{\bf k}', \downarrow} \hat c_{{\bf k}', \uparrow} ,
      \hat c_{{\bf k},\uparrow}^\dag ]  .
\end{eqnarray*}
Here, the Liouville operator $\tilde{\sf L}_t$ corresponds to the commutator with the hopping Hamiltonian 
$\tilde{\cal H}_t= \sum_{\bf k} \tilde{\varepsilon}_{\bf k}\, \hat c_{{\bf k}\sigma}^\dagger
\hat c_{{\bf k}\sigma}$, which agrees with the fully renormalized Hamiltonian 
$\tilde{\cal H}$ in the normal state investigated in paper~I. Therefore, we can use
$ \tilde{\sf L}_t \hat c_{{\bf k},\uparrow}^\dag = \tilde{\varepsilon}_{\bf k}\, 
\hat c_{{\bf k}\sigma}^\dagger$ and find using the anti-commutator relation \eqref{3}
\begin{eqnarray}
\label{C5}
\tilde{\sf L} \hat c_{{\bf k},\uparrow}^\dag &=& \tilde
\varepsilon_{\bf k} \hat c_{{\bf k},\uparrow}^\dag 
- \frac{1}{\sqrt{N}} \sum_{i \ne j} \tilde \Delta_{i,j}^* 
\left(e^{-i {\bf k}{\bf R}_j} {\cal D}_\uparrow (j)\, \hat c_{i,\downarrow} 
- e^{-i {\bf k}{\bf R}_i} S_i^-  \hat c_{j,\uparrow} \right) \, .
\end{eqnarray}
The quantity  $\tilde \Delta_{i,j}^*$
is defined by $\tilde \Delta_{i,j}^* = \frac{1}{N} \sum_{\bf k} \tilde \Delta_{\bf k}^*
  e^{i{\bf k}({\bf R}_i - {\bf R}_j)}$, and ${\cal D}_\sigma(j) = 1- n_{j,-\sigma} = 
{\cal P}_0 + \hat n_{i\sigma}$ was already given in Eq.~\eqref{3}.
The main contribution to the second term in Eq.~\eqref{C5} is caused by 
the following process: First, two holes are generated at sites $i$ and $j$  before 
the hole at $j$ is annihilated again by a local creation operator in
$\hat c_{{\bf k}\uparrow}^\dagger$. The arising local projector ${\cal D}_\sigma(i)$
will be approximated by its average $D= \langle {\cal D}_\sigma(j) \rangle =
1 - \langle n_{j, -\sigma} \rangle$. Thus, we obtain
\begin{eqnarray}
\label{C6}
\tilde{\sf L} \hat c_{{\bf k},\uparrow}^\dag &=& \tilde
\varepsilon_{\bf k} \hat c_{{\bf k},\uparrow}^\dag - \frac{D}{\sqrt{N}}
\sum_{i \ne j} \tilde \Delta_{i,j}^* e^{-i {\bf k}{\bf R}_j} 
\hat c_{i,\downarrow} \nonumber \\
&=& \tilde \varepsilon_{\bf k} \hat c_{{\bf k},\uparrow}^\dag - D \tilde \Delta_{\bf
  k}^* c_{-{\bf k},\downarrow} \, ,
\end{eqnarray}
where $\tilde{\Delta}_{i,j}^*$ was Fourier back transformed to $\tilde{\Delta}_{\bf k}^*$. 
A corresponding contribution from the last term in Eq.~\eqref{C5} vanishes, since
$\langle S_i^- \rangle =0$ outside the antiferromagnetic regime.  
The evaluation of the second commutator in Eq.~\eqref{C4} can be done in 
analogy to the result \eqref{C6}, 
\begin{eqnarray}
\label{C7}
\tilde{\sf L}\hat c_{-{\bf k},\downarrow} &=& -\tilde
  \varepsilon_{\bf k} \hat c_{-{\bf k},\downarrow} - D \tilde \Delta_{\bf
  k} c_{{\bf k},\uparrow}^\dag.
\end{eqnarray}
Inserting  Eqs.~\eqref{C6} and \eqref{C7} into Eq.~\eqref{C4} leads to 
the following two equations for 
${\sf U}_{\bf k}$ and ${\sf V}_{\bf k}$:
\begin{eqnarray}
\label{C8}
{\sf U}_{\bf k} \left( \tilde \varepsilon_{\bf k} - E_{\bf k} \right) + {\sf V}_{\bf k} D
  \tilde \Delta_{\bf k} &=& 0 , \nonumber \\
-{\sf U}_{\bf k} D \tilde \Delta_{\bf k}^* + {\sf V}_{\bf k} \left( \tilde \varepsilon_{\bf
  k} + E_{\bf k} \right) &=& 0  .
\end{eqnarray}
The eigenvalue $E_{\bf k}$ for this system of equations is easily obtained:
   \begin{eqnarray}
 \label{C9}
E_{\bf k}= \sqrt{\tilde{\varepsilon}_{\bf k}^2 +D^2 |\tilde{\Delta}_{\bf k}|^2}.
\end{eqnarray}
The expectation value $\langle \hat c_{{\bf k}, \uparrow}^\dagger
\hat c_{-{\bf k}, \downarrow} \rangle_{\tilde{\cal H}}$, formed with the superconducting
Hamiltonian $\tilde{\cal H}$, is found by solving  \eqref{C2} for 
$\hat c_{{\bf k}, \uparrow}^\dagger$ and $\hat c_{-{\bf k}, \downarrow}$. 
Using the property \eqref{C3}, one finds 
\begin{eqnarray}
\label{C10}
\left\langle \hat c_{{\bf k},\uparrow}^\dag
  \hat c_{-{\bf k},\downarrow}^\dag \right\rangle_{\tilde{\cal H}} &=& \frac{D^2 \tilde \Delta_{\bf
  k}^*}{2 E_{\bf k}} \left( 1 - \frac{2}{1 + e^{\beta E_{\bf k}}} \right).
\end{eqnarray}

\end{appendix}

\end{document}